\documentclass[letterpaper,twocolumn,10pt]{article}
\usepackage{usenix-2020-09}

\usepackage{filecontents}
\usepackage[english]{babel}
\usepackage{blindtext}
\usepackage{enumitem}
\usepackage{subfig}
\usepackage{tikz}
\usepackage{amsmath,algorithm,tabularx}
\usepackage[noend]{algpseudocode}
\usepackage[small,compact]{titlesec}
\usepackage[format=plain,labelfont=bf,font=small]{caption}
\usepackage{titling}
\usepackage{wrapfig}

\newcommand\paragraphb[1]{\noindent{\bf #1.}}
\newcommand\paragraphbq[1]{\noindent{\bf #1}}
\newcommand\paragraphi[1]{\noindent\emph{#1}}

\newcommand{\farmernetes}{Cornet\xspace}

\newcommand{\allnotes}[1]{}
\renewcommand{\allnotes}[1]{#1}

\begin{document}
\setlength{\droptitle}{-3em}

\title{\Large{\textbf{Centralized Management of a Wifi Mesh for Autonomous Farms}} \vspace{-1em}}
\date{}


\author{Ammar Tahir, Yueshen Li, Jianli Jin, Changxin Zhang, Daniel Moon, Aganze Mihigo, \\ Muhammad Taimoor Tariq, Deepak Vasisht, Radhika Mittal\\\\\emph{University of Illinois at Urbana-Champaign}}



\maketitle

\begin{abstract}
    Emerging autonomous farming techniques rely on smart devices such as multi-spectral cameras, collecting fine-grained data, and robots performing tasks such as de-weeding, berry-picking, etc. These techniques require a high throughput network, supporting 10s of Mbps per device at the scale of tens to hundreds of devices in a large farm. We conduct a survey across 12 agronomists to understand these networking requirements of farm workloads and perform extensive measurements of WiFi 6 performance in a farm to identify the challenges in meeting them. Our measurements reveal how network capacity is fundamentally limited in such a setting, with severe degradation in network performance due to crop canopy, and spotlight farm networks as an emerging new problem domain that can benefit from smarter network resource management decisions. To that end, we design \farmernetes, a network for supporting on-farm applications that comprises: (i) a multi-hop mesh of WiFi routers that uses a strategic combination of 2.4GHz and 5GHz bands as informed by our measurements, and (ii) a centralized traffic engineering (TE) system that uses a novel abstraction of resource units to reason about wireless network capacity and make TE decisions (schedule flows, assign flow rates, and select routes and channels). Our evaluation, using testbeds in a farm and trace-driven simulations, shows how \farmernetes achieves 1.4 $\times$ higher network utilization and better meets application demands, compared to standard wireless mesh strategies. 
\end{abstract}

\section{Introduction}

Emerging autonomous farming techniques promise to meet critical goals of increasing farm yields, reducing wastage, and improving agricultural sustainability~\cite{king2017future, schimmelpfennig2016farm, the_economist_2016, vasisht2017farmbeats, fcc-report}. 
For example, bulk spraying of pesticides can be replaced with targeted chemical application at individual affected plants by robots. 
These techniques rely on smart devices such as autonomous robots, drones, and hyperspectral cameras, that require high bandwidth network connectivity for remote operation and data transmission. The widespread adoption of these techniques is fundamentally limited today by the lack of suitable network connectivity on farms. This connectivity challenge has been recognized as one of critical importance by the Federal Communications Commission (FCC) and the US Department of Agriculture (USDA)~\cite{usda_report,fcc-report}. 


In this paper, we study the networking requirements for emerging on-farm applications, and conduct extensive WiFi 6 measurements on a farm to understand the unique challenges in meeting these requirements. Guided by this understanding, we design \farmernetes, a centrally managed WiFi mesh network for farm applications. 


We begin with a survey of 12 agronomists and an extensive literature review to understand the requirements of farm applications. 
Our study, detailed in \S\ref{sec:workload}, highlights the need for a high bandwidth network to support tens to hundreds of devices that include: (i) sensors transmitting large amounts of data to edge servers located in the farmers home/office near the farm (e.g. RGB, thermal and hyperspectral cameras that can generate 10s to 100s of MBs of data per acre, with a typical farm spanning a few hundred acres), and (ii) robots performing targeted bug-spraying, berry-picking, etc, that are remotely controlled by edge servers, with each device requiring a steady upstream of 5-20Mbps. In many cases, these devices must operate from under the crop canopy where they are completely surrounded by crops. 

Keeping the above requirements in mind, we next set out to investigate how much network throughput can we expect to achieve in a farm. For this, we conduct a first-of-its-kind network throughput measurement for WiFi 6 (at both 2.4 GHz and 5 GHz bands) in an 80 acre corn field under different settings (varying distance, varying height of transmission with respect to crop canopy, etc).~\footnote{Our reasons for choosing WiFi, detailed in \S\ref{sec:background}, include its relatively higher bandwidth, its ready availability and its popularity (especially in rural farms \cite{ojha2015wireless, ubiquity_2020, ayrstone_productivity_2023} where commercial cellular connectivity is not an option).} 
Our measurements reveal the severe impact of crop canopy on WiFi performance. We found that: (i) 5GHz WiFi is unable to work well under-canopy, with the effective throughput degrading to zero beyond a short distance of 40m! In contrast, 2.4GHz WiFi is more robust, sustaining close to 10 Mbps of average throughput even at a distance of 100 meters under-canopy. 
(ii) For transmissions above the crop canopy, 5GHz fared significantly better than 2.4GHz, 
achieving around 80 Mbps of throughput at a distance of 100m (2.5 $\times$ higher than 2.4GHz). 

\begin{figure}
    \centering
    \includegraphics[width=0.4\textwidth]{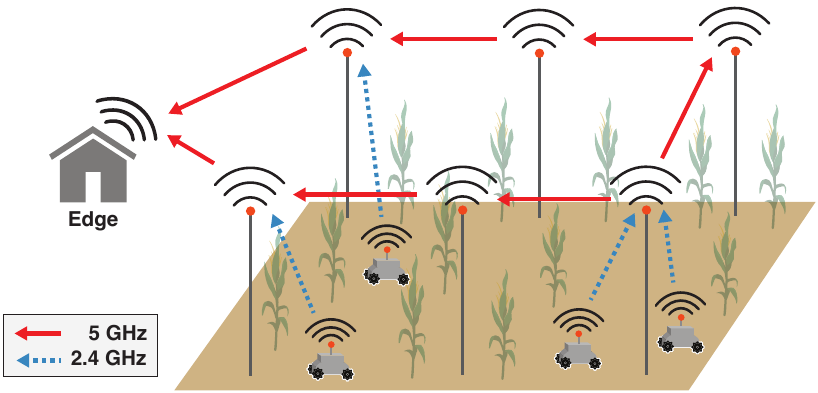}\vspace{-0.1in}
    \caption{\farmernetes' two-tiered architecture uses 2.4 GHz under the crop canopy, and 5GHz above the crop canopy.}\vspace{-0.15in}
    \label{fig:dualplane}
\end{figure}
The insights from our measurements result in a unique WiFi network design for autonomous farms -- a multi-hop mesh of routers placed above the crop canopy (e.g. on poles) that operates in two tiers, as shown in Figure~\ref{fig:dualplane}: (i) \emph{2.4GHz under-canopy tier}, where the routers talk to the under-canopy devices (e.g. robots and sensors) over 2.4GHz channels, and (ii) \emph{5GHz above-canopy tier}, where the traffic is routed to the edge servers across the mesh using the 5GHz band. 

We next come to the challenge of supporting on-farm applications on such a wireless mesh network. This involves supporting tens to hundreds of devices, where each device requires tens of Mbps of bandwidth, and typically operates from under the crop canopy with degraded network throughput. Moreover, multiple devices streaming simultaneously to the edge creates an \emph{incast}, that additionally overwhelms the above-canopy routers closer to the edge. 
Extracting maximal performance under these conditions requires smart management of the limited network resources. 

These challenges bear resemblance to datacenters. Like datacenters, farm networks must support a heterogeneous mix of performance critical and background tasks (remotely controlled robots and sensor data collection). Moreover, unlike typical WiFi settings in homes and enterprises that have unpredictable traffic, farm workloads are more predictable and can be explicitly scheduled.
We can therefore draw inspiration from datacenter and WAN traffic engineering (TE) \cite{hong2018b4, jain2013b4,hong2013achieving} to design a centralized TE engine for \farmernetes. 

The basis of any TE system is its ability to answer the following question: how much network resources will be consumed by a flow that sends data at rate $X$ bits/s along a route $P$? The ability to answer this question allows a TE system to systematically reason about various resource allocation decisions (when to schedule flows, at what rate they should send data, what routes to pick, etc) so as to best utilize the available network resources. Answering this question is straight-forward in wired networks -- a flow sending at a rate $X$ bits/s along a route $P$ will simply consume $X$ bits/s of the network capacity on each link in the path $P$. However, answering this question is a lot trickier in wireless networks. Unlike wired networks, we cannot reason about network capacity independently for each link. Wireless network capacity, given by the channel bandwidth in MHz at a given device, is shared by all transmissions to and from the device, and by interfering transmissions from other devices that use overlapping channels. Moreover, the amount of network capacity that gets consumed at a device is not just a function of the transmission rate, but also the channel quality and distance (transmissions at a given datarate over longer distances or over a degraded channel consume greater amount of channel bandwidth). 

To systematically reason about these effects, we introduce the abstraction of ``resource units'' that captures the fraction of wireless network resources (channel bandwidth) that gets used up by a flow sending at a specified datarate, both at the devices along the flow's path and at the neighboring devices with which it interferes. For example, it allows us to compute that a transmission from device $A$ to router $B$ at rate $X$ bits/s and distance $y$ meters will consume 50\% of the resource units at $A$ and at $B$, and, due to interference, will consume $\approx$25\% resource units at another device $C$ which is at a distance of $z$ meters from $A$. We use this abstraction to reason about different TE decisions -- temporal scheduling of flows, flow rate assignment, channel allocation, and route selection. Our TE decisions also effectively capture network sharing policies (e.g. prioritizing real-time flows over data collection tasks), which are otherwise difficult to realize in wireless networks due to the effects of interference.  
 While we motivate and design our TE system in the context of farm networks, it is broadly applicable to other contexts where wireless mesh networks are used to support predictable and schedulable workloads (e.g. factories, smart cities, mining, etc). 

We use a combination of controlled on-field experiments and trace-driven simulations to evaluate \farmernetes.
Our evaluation shows how \farmernetes results in upto 1.4 $\times$ better network utilization and  2.5-4 $\times$ better performance for real-time tasks (in terms of throughput normalized by flow demand) when compared with standard wireless mesh strategies. We summarize our contributions below: 

\noindent $\bullet$ We conduct the first-of-its-kind measurements of WiFi throughput in a cornfield. 

\noindent $\bullet$ Guided by these measurements, we design \farmernetes, a two-tiered network for autonomous farms. 

\noindent $\bullet$ We develop a centralized traffic engineering engine for \farmernetes, that uses a novel resource unit abstraction to reason about wireless network capacity.

\noindent $\bullet$ We evaluate \farmernetes using trace-driven simulations and real-world testbeds. 

\section{Requirements and Challenges}
\label{sec:challenges}

\subsection{Understanding Farm Workloads}
\label{sec:workload} 


We surveyed 12 agronomists spanning two institutions, who are conducting cutting-edge research in autonomous farming techniques.  We used the survey responses, in combination with an extensive literature review of precision agriculture \cite{king2017future, schimmelpfennig2016farm, the_economist_2016, vasisht2017farmbeats, fcc-report} and autonomous farming 
 techniques~\cite{bogue2020fruit, kamtikar2021fruitarm, tang2020fruitrecognition, sivakumar2021visualnav, insect_detection, barnes2021opportunities, billingsley2008robotics, rose2021responsible}, to characterize networking requirements for emerging farm workloads.

\paragraphb{Workload Categories} We divide networked workloads for farms into two categories:

\paragraphi{(a) Real-time Tasks.} These tasks include targeted bug-spraying \cite{insect_detection}, fruit picking \cite{bogue2020fruit, kamtikar2021fruitarm, tang2020fruitrecognition}, mechanical weeding \cite{li2022key, ziwen2015study}, and cover crop planting \cite{barnes2021opportunities} by autonomous under-canopy robots. While tasks like visual navigation (e.g. moving along the crop row) can be executed entirely on the robot device, autonomous plant manipulation tasks (such as the ones exemplified above) rely on more complex machine learning models, that are computationally too heavy to run on a power-constrained robot device. These tasks, therefore, need to leverage edge servers to run complex models for controlling the robots. 
This requires streaming sensor data (e.g. RGB images from multiple cameras) in real-time from under-canopy robots to edge servers, with the throughput requirements consequently ranging from 5-20Mbps per device. Visual navigation via tele-operated robots and drones~\cite{sivakumar2021visualnav} similarly requires an upstream throughput of 5-10Mbps.  
At scale, a typical farm spanning a few hundred acres may require many tens of robots to operate simultaneously. 



\paragraphi{(d) Data collection.} Standard agricultural sensors, e.g. to monitor pH and water levels, send small amounts of data (few 10s of KBs). 
Emerging precision agriculture techniques additionally rely on sensors such as RGB, thermal, and multispectral cameras that 
generate much larger amounts of data (10s to 100s of MBs per acre).
In many cases, such data must be collected multiple times a day, to catch potential issues faster and take appropriate measures. There can be tens to hundreds of such data collection devices in a large farm.

\paragraphb{Survey Takeaways} Our survey participants are actively engaged in research spanning the two application categories described above. Our survey asked pointed questions about their networking needs. We summarize the takeaways below:

\noindent $\bullet$  All participants (N=12/12) feel that a high throughput network will make it easier to deploy smart agriculture solutions and enable innovation in the space.

    
\noindent $\bullet$  The requirements of our participants included network support for real-time tasks (N=4/12), e.g. for remotely controlling autonomous robots, and processing data collected by on-field sensors within a few minutes (N=1/12), a few hours (N=3/12), or a few days (N=3/12). 

\noindent $\bullet$  Data collection applications use various sensors (RGB, thermal, hyperspectral, LIDAR, thermometer, pH, and spectroradiometers). RGB images were most common (N= 11/12), followed by hyperspectral (N=6/12), and LIDAR (N=4/12).

\noindent $\bullet$  Majority of our participants required processing the sensed data in edge servers (N=8/12), with a few indicating need for cloud offloads (N=3/12). 

\paragraphb{Key Workload Characteristics} We summarize key characteristics of networked farm workloads.

\noindent $\bullet$ \emph{Need for high bandwidth.} 
In order to support tens of simultaneous real-time data-streams and upload large amounts of data from multiple sources within a few hours, the network infrastructure in the farm must ideally support a capacity of a few hundreds of Mbps. 

\noindent $\bullet$ \emph{Under-canopy operation.} Many applications (e.g. bug-spraying, cover-crop planting, berry-picking, etc) involve streaming data from under the canopy, where the transmitting device is completely surrounded by crops, leading to degraded network performance. 

\noindent $\bullet$ \emph{Incast to edge.} A typical farm workload would comprise of multiple on-field devices  simultaneously streaming data to the edge servers (typically located in farmer's home/office or in service trucks beyond the edge of the field). 

\noindent $\bullet$ \emph{Heterogeneity and flexibility.} Farm workloads allow room for scheduling in two ways: (i) Applications vary in their requirements and can be prioritized accordingly, e.g. real-time tasks can be prioritized over data collection tasks that have laxer deadlines. (ii) Most tasks have some flexibility in when they must be started. Even real-time tasks can be started a few minutes after they have been requested. Some tasks may further be pre-emptible, thereby allowing greater flexibility. For example, data collection or even a real-time task such as fruit-picking can be paused to favor a more urgent task (e.g. tele-operation by human experts). 


\subsection{The Ag-Connectivity Challenge}\label{sec:background}

Enabling reliable high-bandwidth connectivity for agricultural applications at scale in large farms is challenging because of limited availability of infrastructure. 
Sensors and networked devices cannot be plugged into power sockets or connected to Ethernet backhauls, unlike traditional enterprise networks. Furthermore, due to the sparse population of farmlands, there is limited deployment of infrastructure-heavy solutions like cellular and fiber networks. Recently, government bodies, such as FCC~\cite{fcc-report} and US Department of Agriculture~\cite{usda_report}, have recognized this challenge and invested heavily in establishing fiber connectivity in rural areas. However, even where such fiber endpoints are available, they can only serve as backhauls and must be extended to devices on the field through wireless connectivity. We discuss some options for on-farm wireless connectivity below:

\noindent\textbf{(i) Low power wide area networks (LPWANs):} LPWANs, such as LoRa~\cite{lora}, are one of the most popular solutions for long range low power connectivity. They support a range of up to 10 Kilometers in rural areas, but provide only a few Kbps bandwidth, failing to meet the high bandwidth requirements of emerging farm applications (\S\ref{sec:workload}). 

\noindent\textbf{(ii) Satellite Networks:} Satellite networks, such as Starlink, promise to provide universal high-bandwidth connectivity. 
However, they are not suitable for on-farm operation because: (a) satellites use high-frequency signals (8-10 GHz) which are easily blocked by crops,  (b) the satellite network terminals are relatively large in size and cannot be carried by farm robots.

\noindent\textbf{(iii) TV White Spaces (TVWS):} TVWS are empty parts of the TV spectrum used to transmit data signals~\cite{whitefi}. TVWS offers long range (10-15 Km) and relatively high bandwidths (tens of Mbps). However, to make use of TVWS, 
the devices must be fitted with bulky and power-hungry antennas. The amount of available bandwidth may also vary over time based on how much of the spectrum is unoccupied.  


\noindent\textbf{(iv) Citizen Band Radio Service (CBRS):} CBRS is an emerging technology that operates in lightly-licensed spectrum around 3.5 GHz. CBRS allows users to setup their own cellular-like networks without requiring spectrum purchase. CBRS is a promising solution for on-farm connectivity, due to its long range (few miles) and cellular-like bandwidths. However, we expect that it would experience similar challenges for under-canopy operation as discussed in \S\ref{sec:measurements} due to its relatively high frequency. In addition, it would require sensors and robots to be fitted with (power-hungry) cellular-connectivity options. Nonetheless, CBRS could be a viable connectivity option for farms, that can co-exist with WiFi (which offers several advantages, as discussed below).

\noindent\textbf{(v) WiFi:} WiFi is already integrated into most sensors, robots, and cameras, which makes WiFi networks plug and play. WiFi is also globally available (e.g. as opposed to CBRS and TVWS, that are available only in a few developed countries~\cite{cbrs-countries, tvws-countries}). Our measurements in \S\ref{sec:measurements} show how WiFi can support the required bandwidths of up to 10s to 100s of Mbps, but it has limited range (tens of meters). It must therefore operate as a mesh network for on-farm applications.
The relatively limited range of WiFi also has some benefits. It would consume lower power on sensors and robots than longer range alternatives like CBRS, and allow spatial re-use of the available spectrum. In contrast, the shared long-range CBRS spectrum may suffer greater interference, not only from simultaneous transmissions on the same farm, but also from neighboring farms. Wi-Fi also has more overall spectrum (over 400MHz across different bands) compared to CBRS (150 MHz shared with higher priority users, with only 20-40MHz reserved for general use~\cite{cbrs-band}). Therefore, a WiFi mesh network can potentially offer higher and more predictable bandwidth.





In this paper, we focus on using WiFi due to these relative advantages. However, designing an on-farm WiFi mesh network that can meet the performance requirements of emerging farm applications requires tackling multiple challenges:
How does crop canopy impact WiFi performance? How do we strategically use 2.4GHz and 5GHz bands in our WiFi mesh given their different performance characteristics, particularly in presence of crop canopy? What forms of performance bottlenecks emerge in such a network? How do we design a centralized TE system that can leverage the predictability and schedulability of farm workloads in order to alleviate the network bottlenecks and extract better performance? How do we reason about wireless capacity (capturing the effects of channel quality, transmission distance, and interference) in order to make TE decisions? What are the specific TE decisions that we must make and what algorithms should we develop for them? How do we enforce different policies, e.g. ensuring that background data-collection tasks only use the ``spare'' network capacity left unused by the real-time tasks (including at the devices where they interfere)? The remainder of our paper addresses these questions. 

\section{On-Farm WiFi Measurements}
\label{sec:measurements}

\begin{figure}[t]
\centering
\captionsetup{justification=centering}
\captionsetup[subfigure]{justification=centering}
\centering
\subfloat[Receiver]{\includegraphics[width=0.13\textwidth]{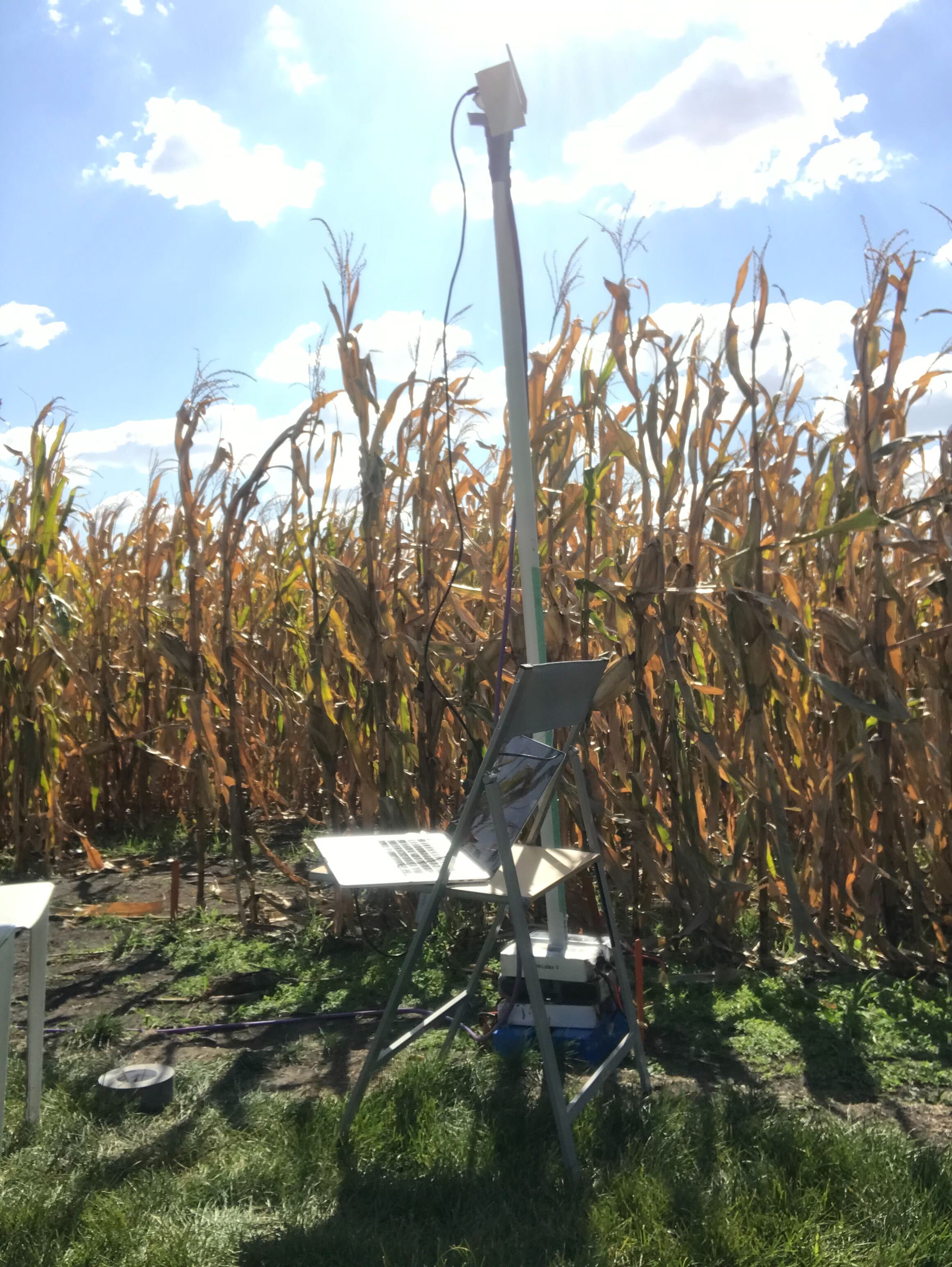} \label{fig:receiver}}
\subfloat[Sender Under Canopy]{\includegraphics[width=0.13\textwidth]{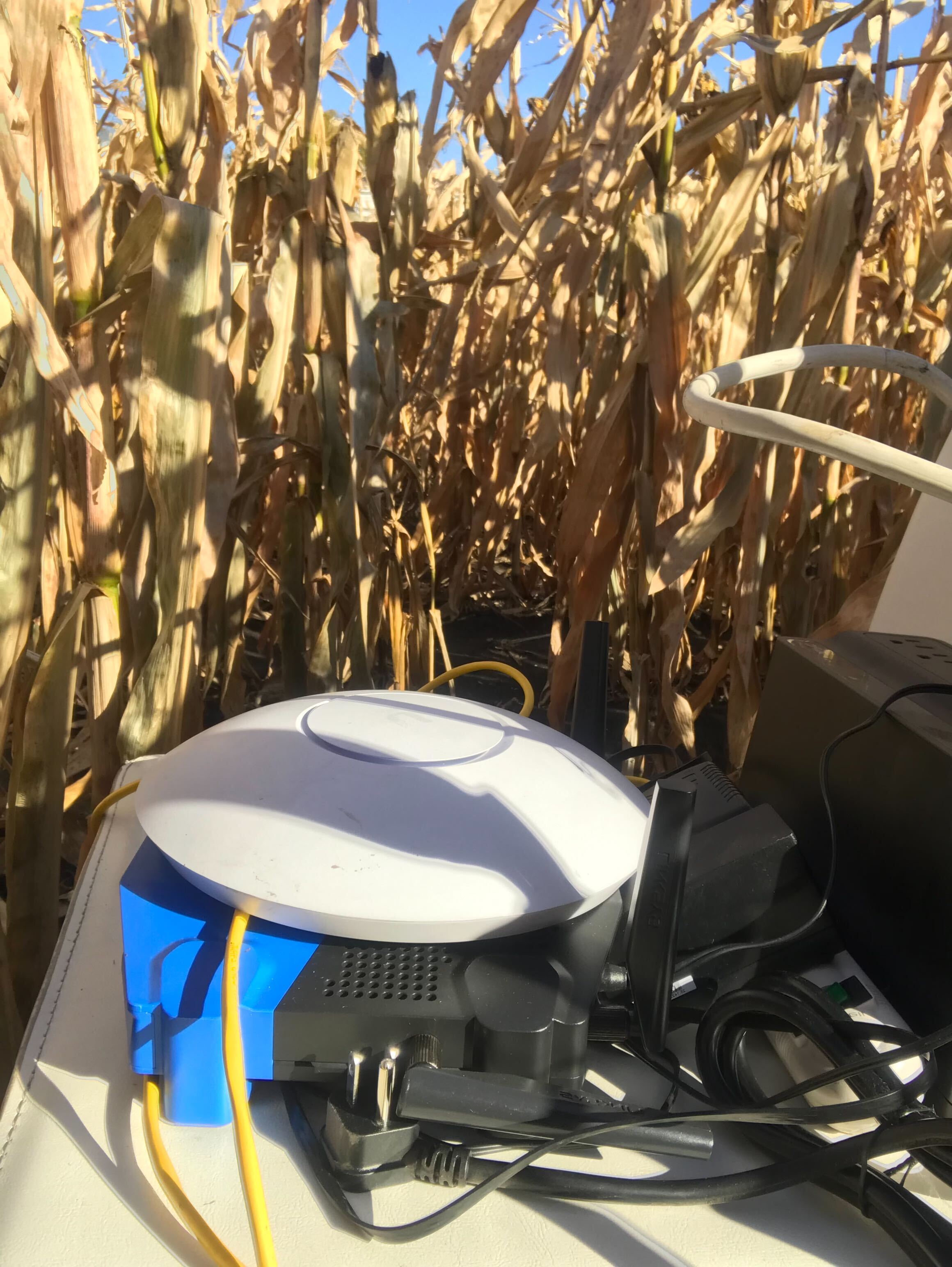} \label{fig:sender-uc}}
\subfloat[Sender Above-Canopy]{\includegraphics[width=0.13\textwidth]{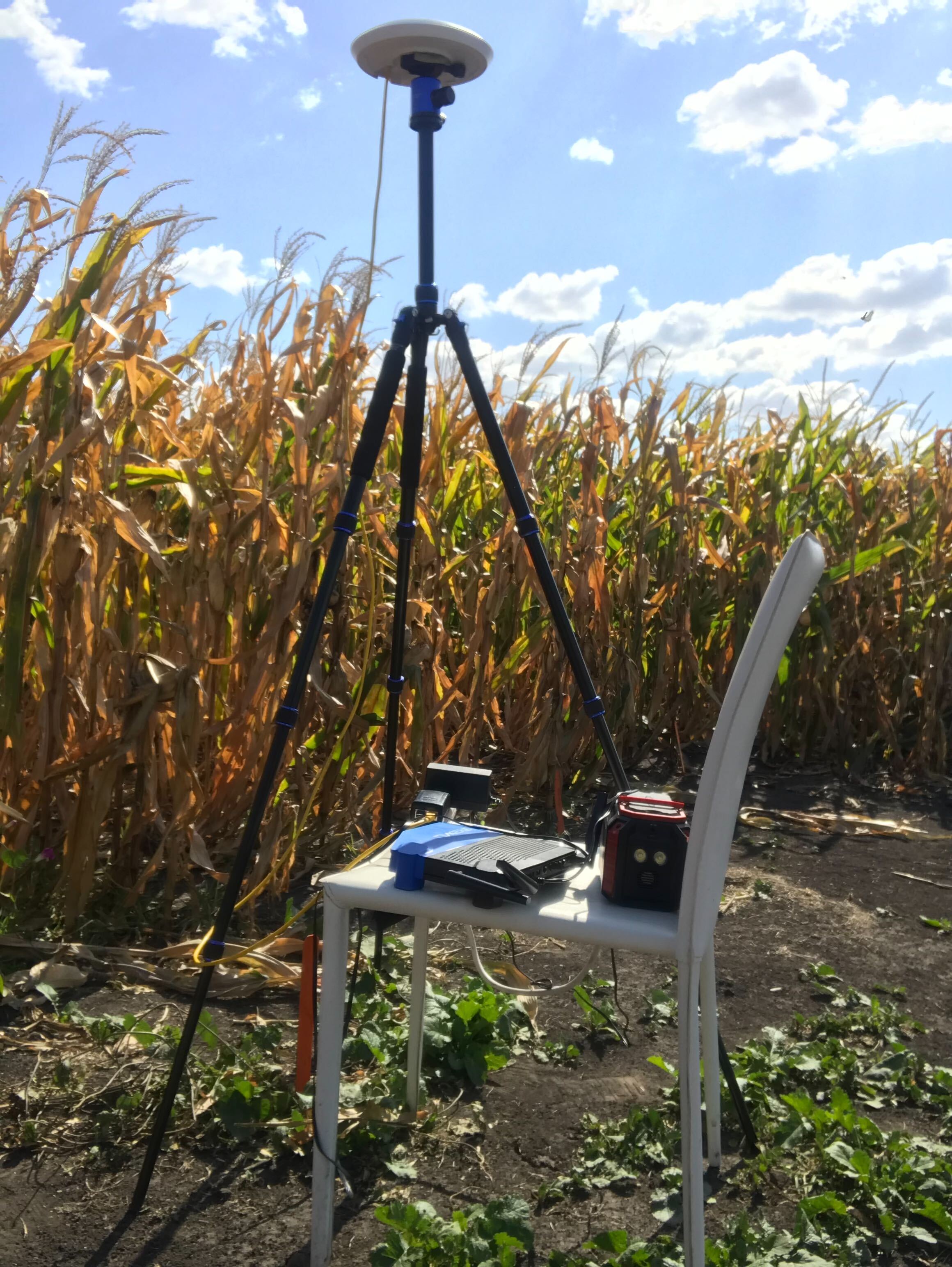} \label{fig:sender-ac}}\\
\vspace{-10pt}
\caption{Setup: Receiver is a router placed on a pole above canopy, Sender is tested with both above canopy and under canopy placement.}
\vspace{-10pt}
\label{fig:setup}
\end{figure}

\begin{figure*}[t]
\centering
\captionsetup{justification=centering}
\captionsetup[subfigure]{justification=centering}
\centering
\subfloat[Wifi over 2.4 GHz]{\includegraphics[width=0.31\textwidth]{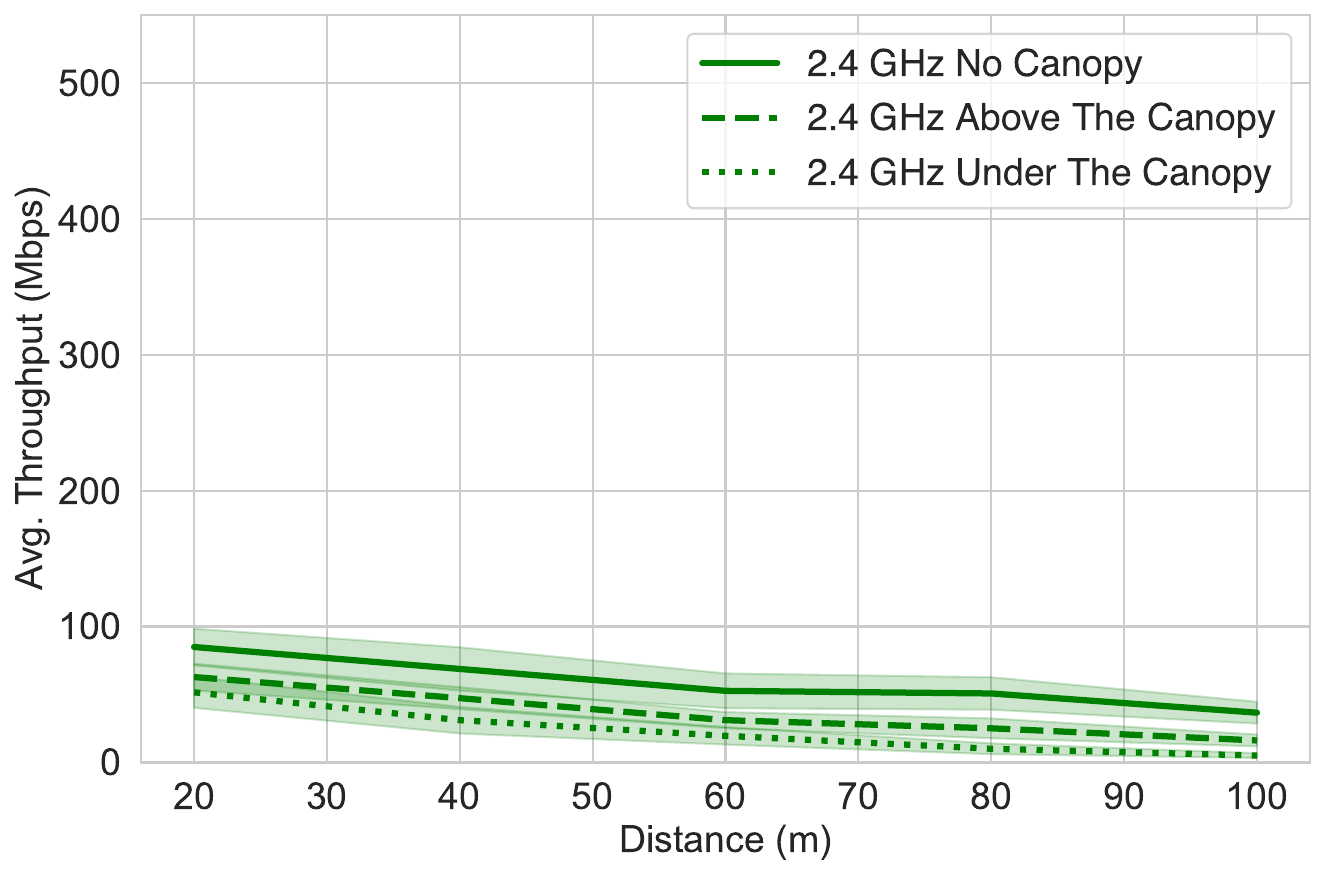} \label{fig:2_4ghz1}}
\subfloat[Wifi over 5 GHz]{\includegraphics[width=0.31\textwidth]{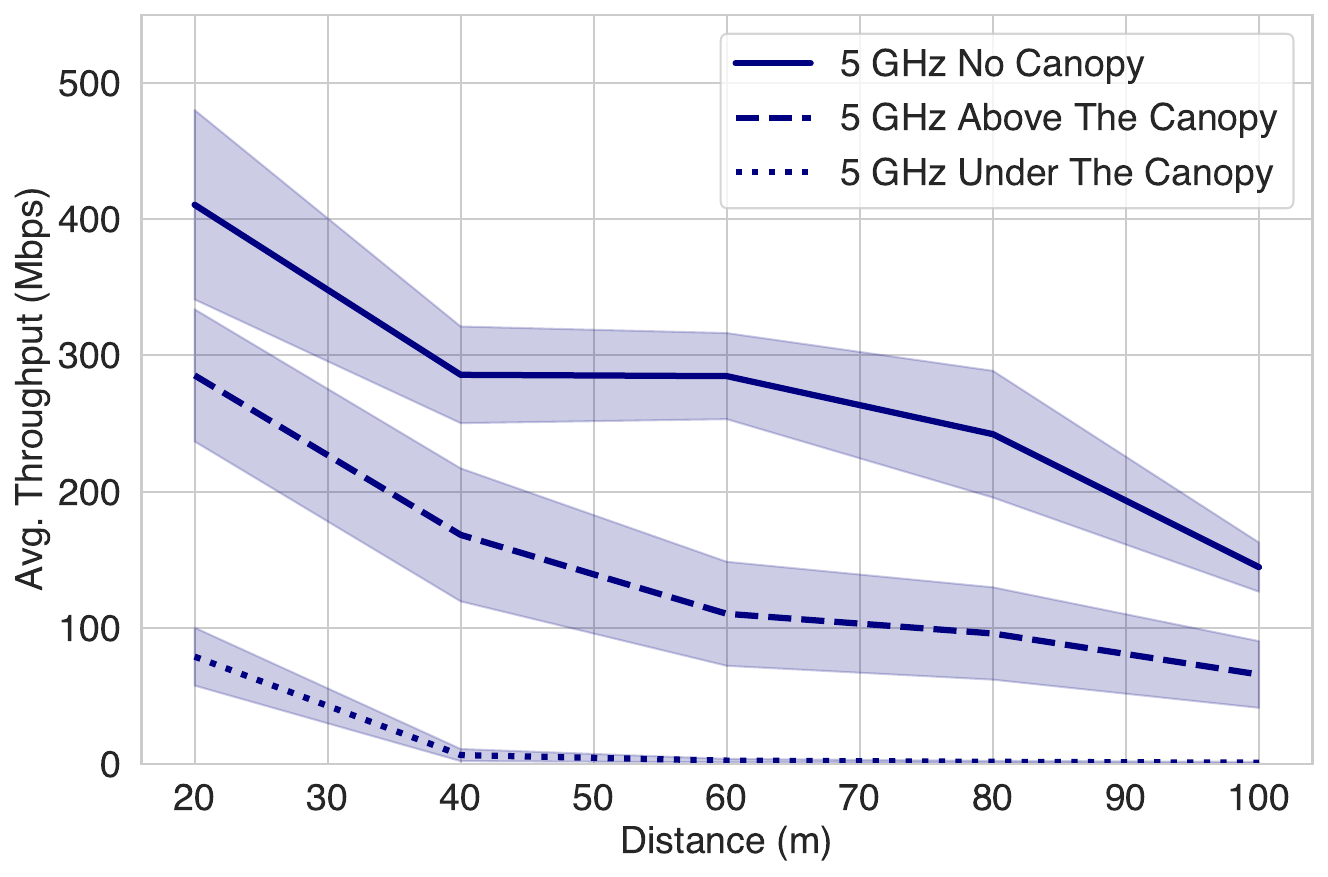} \label{fig:5ghz1}}
\subfloat[Comparison of 2.4 GHz and 5GHz UC]{\includegraphics[width=0.31\textwidth]{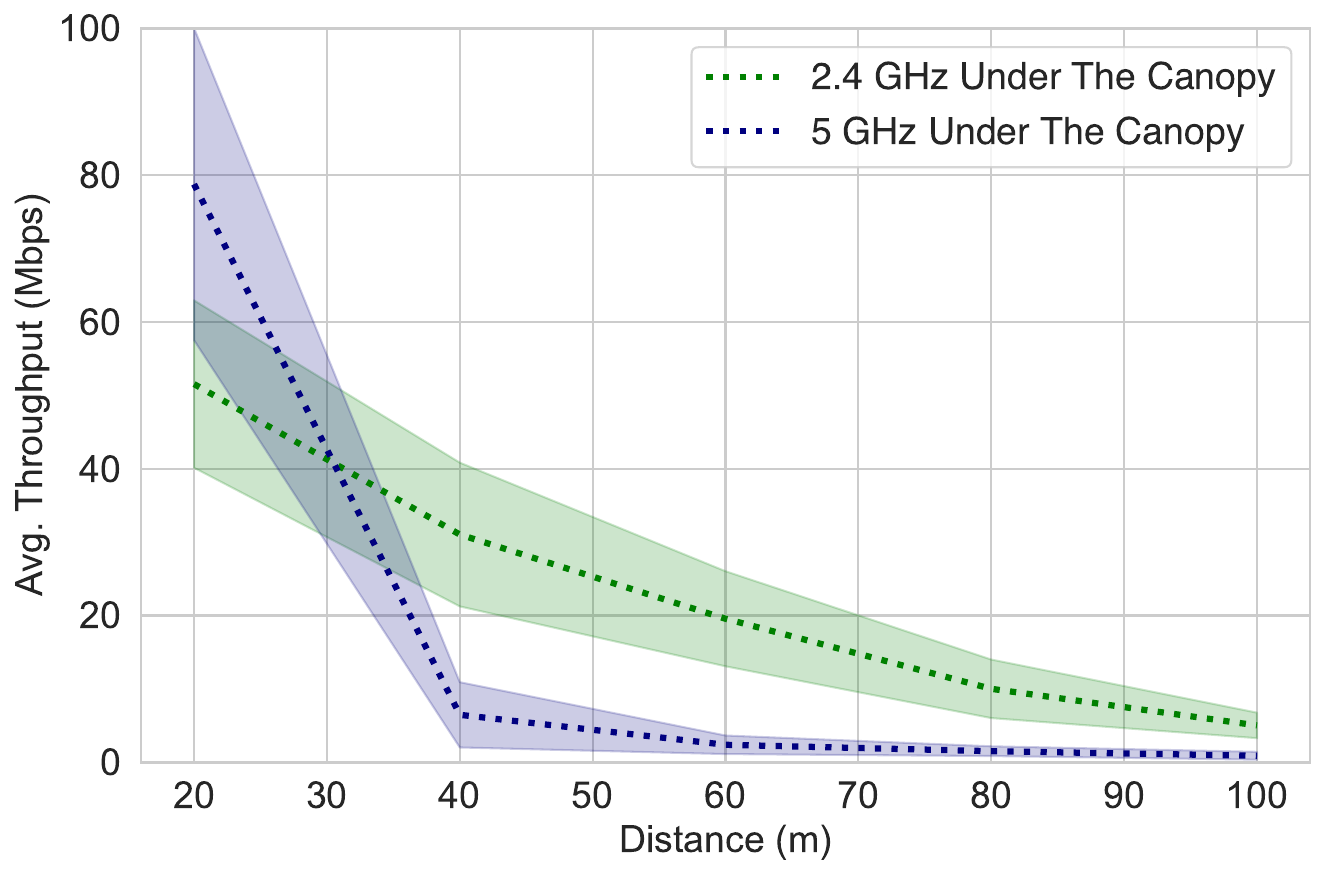} \label{fig:undercanopy}}
\vspace{-10pt}
\caption{Wifi throughput with increasing distance (shaded area shows standard deviations across different experiment runs over the growing season).}
\vspace{-10pt}
\label{fig:distance1}
\end{figure*}

Farms provide a unique environment to study WiFi performance -- unlike typical enterprise settings where WiFi is used, farms do not suffer from interference from external sources, allowing use of most of the WiFi spectrum. However, network throughput and range can be impacted by the presence of crop canopy. We study this by measuring WiFi performance in a 32-acre cornfield, with about 6 feet high crops.

\subsection{Experiment Setup}

Our setup, described below (and shown in Fig \ref{fig:setup}) uses state-of-the-art WiFi 6 devices. 

\paragraphb{Sender} We used a Unifi LR-6 router that supports Wifi 6 over 5GHz and 2.4GHz bands and is designed for outdoor environments \cite{unifilr6}. We used a 4 Gigabit LAN port to connect the router to a 1.8 GHz dual-core Linux device that sends a UDP flow using iperf. 

\paragraphb{Receiver} We used a TP-Link RE505X receiver. Our receiver supports Wifi 6 over 5GHz and 2.4GHz bands. The receiver obtains data from the sender-side Unifi LR-6 router over WiFi. We used Ethernet to connect the receiver to an 8-core 2.4 GHz device running macOS that ran the iperf server to receive the transmitted traffic and log the throughput.~\footnote{We did not observe significant differences in propagation delay (ping latency), which remained roughly 2ms across different experiment settings. We therefore focus on measuring network throughput in our study.}

\paragraphb{Channel configuration} We conducted our experiments using both the 2.4GHz and the 5GHz bands. To evaluate best-case throughput, we maxed out the available bandwidth in each band by doing channel bonding i.e. we used a 160 MHz channel for the 5GHz band and a 40 MHz channel for the 2.4GHz band. We avoided the effect of potential interference by scanning all WiFi networks in the location before each experiment and only picking unused channels for the experiment.

\paragraphb{Transmission Height} We placed the receiver on a 2 m high pole (which was higher than the crop canopy level), as shown in Figure \ref{fig:receiver}. We placed the sender at two different heights to capture different scenarios across experiments: 

\paragraphi{Under-canopy (<0.5m):} This setup (as shown in figure \ref{fig:sender-uc}) mimics robots streaming sensor data from under the crop canopy. Farm robots typically have a short height of about 0.5m for easier maneuverability amidst crops. It is unrealistic to place a WiFi antenna over the canopy height for each robot as it can damage the crops and hinder robot's motion and ability to perform tasks.

\paragraphi{Above-canopy (2m):} This setup mimics settings where sensors and relay routers are mounted on poles above the canopy's height (figure \ref{fig:sender-ac}). 

We also conduct control experiments on a path where there is no crop canopy between the sender and the receiver. We use a transmission height of 2m for these experiments. 

\paragraphb{Transmission Distance} We vary the distance between the WiFi sender and receiver across experiments, with the values ranging from 20m to 100m.  

\paragraphb{Experiment Duration} For each setting, we conducted 5 experiment runs. Each run involved transmitting data for 30 seconds and recording the resulting throughput. We manually moved the sender by few centimeters between different runs for a given setting, to capture performance deviations arising from finer-grained spatial variations. We repeated these measurements every week starting from mid-June up until the end of August for the 2023 season and the month of October for the 2022 season. In total, we collected measurements over 12 weeks accounting for over 1000 individual tests amounting to 800 minutes.

\subsection{Overall Results}
\label{sec:distance-throughput}





Figure~\ref{fig:distance1}(a) and (b) show the average throughput (aggregated across all the measurements we conducted over the growing seasons) achieved by WiFi over the 2.4GHz band and 5GHz bands under different settings -- no canopy (baseline), above canopy and under canopy.  The shaded areas represents the standard deviation over different experiment runs. 

\paragraphb{Takeaways}  

\paragraphi{(i) Presence of crop canopy degrades throughput.} 
Even when the sender and the receiver are in line of sight above the crop canopy, the WiFi throughput achieved on either bands is 1.2-2 $\times$ lower than when there is no crop canopy.~\footnote{This is potentially caused by the absorption of radio waves by crops leading to worse propagation characteristics~\cite{zigbee-undercanopy}.} 

\paragraphi{(ii) 5GHz achieves higher throughput than above canopy.} This is due to wider 160MHz channel with 5GHz as compared to 40MHz wide channel with 2.4GHz. 

\paragraphi{(iii) 2.4GHz is more robust under canopy.} Figure \ref{fig:undercanopy} re-plots the data for the under-canopy setting, for a more direct comparison between the 2.4GHz and 5GHz bands. 
The throughput achieved by 5GHz drops to nearly zero beyond 40m. On the other hand, the 2.4GHz band can give a decent throughput of 5-10Mbps even at a distance of 80 meters. While one would expect 2.4GHz to be more robust to obstacles, we were surprised by the severity with which 5GHz throughput degrades under canopy, in spite of using a wider band.  




\paragraphb{Implications} Our results have the following implications on a Wifi based network in farms: (i) WiFi throughput degrades quickly with distance, indicating the need for a multi-hop mesh network to provide sufficient coverage. (ii) The mesh network should place relay routers on poles at a height above the crop canopy, and use the 5GHz band for above-canopy transmissions. (iii) Under-canopy transmissions would be restricted to using 2.4GHz WiFi (as discussed in \S\ref{sec:two-tier}).

\subsection{Variations over growing season, time, and space}
\label{sec:crop-age}

We report the complete analysis of how throughput varies over weeks, or over small time frames i.e. during a single measurement, and due to small changes in the position of devices in Appendix \ref{appendix:more_measurements}. Here we summarize the key takeaways:

\paragraphb{Trends over the growing season} We find that the overall trends from \S\ref{sec:distance-throughput} remained consistent over weeks during the growing season, with occasional variations in absolute throughput values, owing to factors such as weather and crop height. 

\paragraphb{Temporal Variations} We notice very small temporal variation (10\%) over the duration of an experiment run on a stationary device. 

\paragraphb{Spatial Variations} Changing the sender's position by even a few centimeters results in up to 30 \% standard deviation in throughput. We model this variation in our evaluation in \S\ref{sec:eval}.

\section{\farmernetes' Two-Tiered Architecture}
\label{sec:two-tier}


Guided by our measurement results, \farmernetes architecture comprises of a multi-hop mesh of WiFi routers placed above the crop canopy (e.g. mounted on poles) in a grid topology (as shown in Figure \ref{fig:dualplane}). These routers operate in two distinct tiers : \emph{(1) 2.4GHz under-canopy:} where the
 routers communicate with under-canopy robots and sensors over the 2.4GHz interface, and
\emph{(2) 5GHz above-canopy:} where the routers use the 5GHz interface to route traffic across the mesh above the crop canopy in order to communicate with the edge servers (or with other on-field devices).

This two-tier architecture can be supported using commodity WiFi routers that are typically dual-band, having one interface each for 2.4GHz and 5GHz. The 2.4 GHz interface operates in Infrastructure mode and acts as Access Point (AP) for robots whereas the 5GHz interface operates in ad-hoc mode to form a mesh.

\paragraphb{Router Density} We expect routers to be placed in a grid at a distance of 80-100 m. Increasing the router density produces limited boost in network performance (given that wireless capacity is limited by the shared spectrum bandwidth), and is not worth the increase in cost of deployment. 

\paragraphbq{Why the two tiers?} 5GHz does not work under-canopy beyond 40m, restricting majority of under-canopy transmissions to 2.4GHz. But two questions remain: (i) Could we have used 2.4GHz above canopy to augment 5GHz throughput? Doing so might also allow two neighboring 5GHz routers to operate on different channels (to minimize interference) with the 2.4GHz interface acting as a bridge for sending flows between them. We considered adopting this idea, but dismissed it when our evaluations (\S \ref{sec:trace-eval}) revealed that the longer range of 2.4GHz above canopy makes it ill-suited for routing in the mesh, and severely impacts the under-canopy transmissions (that would then invariably interfere with the longer range above canopy 2.4GHz transmissions). (ii) A related question is why not use 5GHz under-canopy when feasible (for transmissions within a distance of 40m). We found that doing so again results in degraded performance. This is because using the 5GHz interface for the under-canopy transmission would then requires using the 2.4GHz interface for forwarding the above-canopy traffic, which leads to increased interference for the reasons mentioned above. 

Despite our strategic use of 2.4GHz and 5GHz WiFi bands, network capacity is still limited in a farm setting and performance bottlenecks can emerge in both tiers: 

\paragraphb{Network bottlenecks in 2.4GHz tier} Even though 2.4GHz performs better than 5GHz under-canopy, it has limited channel bandwidth and suffers from degraded channel quality due to the crop canopy (achieving a throughput of 10 Mbps over a distance of 80 m). Moreover, the 2.4GHz band only supports three non overlapping channels. So while neighboring routers can be configured with different 2.4GHz channels to minimize interference for under-canopy transmissions, with a choice of only three channels, contention cannot be eliminated when there are four or more active devices in each-other's range.

\paragraphb{Network bottlenecks in 5GHz tier} The 5GHz interface must operate over the same channel at each router to effectively route flows across the mesh. ~\footnote{All routers must operate on the same channel in the 5 GHz tier to avoid network partition.}  Bottlenecks emerge in the 5GHz tier when multiple devices simultaneously stream data to the edge server, creating an incast and overwhelming the routers closest to the edge. The interference caused by neighboring routers (all operating on the same channel) further aggravates the issue. As per our measurements, the throughput for above canopy 5GHz transmissions is limited to 100Mbps at 80m to begin with. Combining the effects of interference, fewer than 10 devices simultaneously transmitting to the edge at 10Mbps would end up overwhelming the network capacity.~\footnote{Using tri-band routers that support two 5GHz interfaces enables using an additional channel in the 5GHz tier, thereby doubling the available bandwidth. Nonetheless, the issue of limited capacity, that is aggravated by incast, still holds, especially as the scale of deploying smart agricultural devices increases.}



\section{\farmernetes' TE System}
\label{sec:te}



To alleviate the above bottlenecks, \farmernetes' incorporates a centralized traffic engineering (TE) engine. 
Inline with the ideas behind software-defined networking, the TE engine runs at a centralized server (residing at the edge) and remotely controls on-field routers and devices.~\footnote{Exchange of control information between the server and the devices has low bandwidth requirements, that can use an out-of-band LoRA network, as opposed to using the in-band WiFi infrastructure.} 


\paragraphb{Inputs} The TE engine takes as input: (i) Locations of on-field devices and expected trajectories of mobile robots. (ii) Network requirements of tasks requested by the farm operator. Requirements for a real-time task include (a) its duration   (b) the deadline by when it should finish, and (c) the \emph{desired} upstream bandwidth for real-time streaming (this can be pre-estimated from the sensing modality, e.g.  RGB vs LiDAR, and the data quality requirements of the downstream data-processing algorithms running at the edge).
Requirements for tasks involving data collection for non-realtime processing include: (a) the deadline by when on-field data must be collected, and (b) the amount of data that must be collected. 

\paragraphb{Knobs} 
\farmernetes invokes the TE engine periodically (every few minutes). At each invocation, 
the TE engine computes the following knobs and remotely configures them on the on-field devices and sensors: (i) the time at which a task (flow) should start, where a task might be paused and resumed by subsequent invocations of the TE logic, (ii) what rate should be assigned to each flow, (iii) which router (AP) should an under-canopy sensor or robot connect to, (iv) channel configurations for the 2.4GHz tier, and (v) routes taken by flows in the 5GHz tier. 

\subsection{Reasoning about Wireless Network Capacity}
\label{sec:resource-unit}


TE decisions require reasoning about the amount of network capacity consumed by a flow. 
We reason about wireless network capacity using the abstraction of resource units, that captures the normalized amount of capacity at each device operating at a given channel. We apply the same reasoning for both the 5GHz and the 2.4GHz bands, although the absolute amount of channel bandwidth differs for the two. A device starts with one resource unit available on each band (which corresponds to 20 MHz of bandwidth for the 2.4GHz band and 160 MHz of bandwidth for the 5GHz band). Direct transmissions to and from this device, and interfering transmissions from other devices, would consume different fractions of resource units on the device for the corresponding band. The fraction of consumed resource units provides a generalized abstraction to represent the share of channel bandwidth consumed by the transmission, where the share can be realized either over time domain (capturing how often the channel is used, e.g. by CSMA based 802.11ac protocol) and/or over frequency domain (capturing how much of the channel is used, e.g. with OFDMA-based wireless technologies \cite{weller2020wifi6ofdma}). 

Our model for computing the fraction of resource units consumed by a flow is guided by data-driven inputs. In particular, we assume we have a trace of WiFi throughput at varying distances (for under-canopy transmission over 2.4GHz and above-canopy transmissions over 5GHz).~\footnote{We can equivalently measure the signal strength at varying distances, and then back-compute maximum achievable throughput from it using known formulations \cite{shannon1949communication}.} We use the traces collected by our measurements described in \S\ref{sec:measurements} -- the trace provides us a measure of throughput at discrete distances, and we apply a logarithmic regression to it to predict throughput for arbitrary distances. In practice, such traces could be collected periodically (say every few weeks). We make a simplifying assumption that for a given tier (2.4GHz under-canopy or 5GHz above-canopy) and at a given point in time, the maximum achievable throughput or signal strength (in the absence of contention or interference from other flows) primarily depends on distance, given relatively static and homogeneous farm setting -- this is corroborated by our observations in \S\ref{sec:measurements}.   


\paragraphb{Resource units consumed due to direct transmission} Consider a flow $f$ from $A$ to $B$ with datarate $X$ Mbps over distance $d_{AB}$, as shown in the figure below. 
\begin{wrapfigure}[5]{r}{0.2\textwidth}
\centering
\vspace{-10pt}
\includegraphics[width=\linewidth]{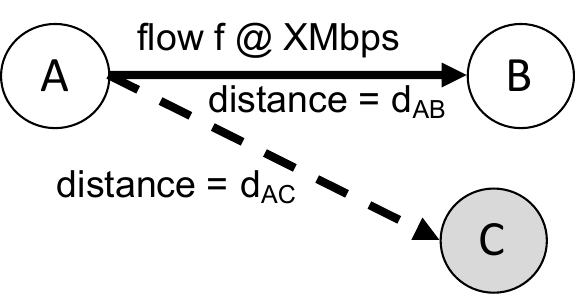} 
\vspace{-10pt}
\end{wrapfigure}
Let $throughput(d_{AB})$ denote the maximum datarate a transmission between $A$ and $B$ could have achieved if it was using the entire channel bandwidth (i.e. one resource unit) at $A$ and at $B$ (this value is available from the measurement traces mentioned above). A transmission at a lower datarate would consume proportionally smaller amount of resource units. Therefore, the amount of resource units (or fraction of channel bandwidth) that gets consumed at $A$ and $B$ by flow $f$ can be denoted by:
 \begin{subequations}
 \begin{align}
    r(f, A) = r(f,B) = \frac{X}{throughput(d_{AB})}\label{eqn:throughput-distance}
 \end{align}
\end{subequations}
where $r(f, A)$ and $r(f,B)$ are resource units consumed by $f$ at $A$ and $B$ respectively. 

\paragraphb{Resource units consumed on interfering devices} We next approximate the amount of resource units consumed by the flow $f$ due to interference on a neighboring device $C$ at a distance of $d_{AC}$ from $A$: 
 \begin{subequations}
 \begin{align}
    r(f, C) =& r(f,A) \times \delta(f,C) \\
    =& r(f,A) \times min(1, \frac{throughput(d_{AC})}{throughput(d_{AB})})
    \label{eqn:interference}
\end{align}
\end{subequations}
Here, the term $r(f,A)$ indicates how often (or how much of) the channel is accessed by flow $f$, and the term $\delta(f,C)$ captures the impact of interference on the portion of channel accessed by $f$.
Our formulation captures the decreasing impact of interference with increasing distance between $A$ and $C$. 

When $d_{AC} \leq d_{AB}$, the interference effect is so high that no part of the spectrum used by $f$ is usable by $C$ (in most 802.11ac implementations, this is realized as the sender backing off when it overhears another transmission). In such a setting, $throughput(d_{AC}) \geq throughput(d_{AB})$, and $r(f, C) = r(f,A)$, which captures the effect that whenever $f$ consumes resources $A$, the corresponding amount of resources (on the shared channel) cannot be used at $C$. 

As $d_{AC}$ increases beyond that, the interference effect materializes as degradation in signal to noise ratio. This means that the portion of channel impacted by $f$ (i.e. $r(f,A)$) is still usable by $C$, but not very effectively. A transmission at a specified datarate to/from $C$ which would have otherwise required a smaller amount of channel bandwidth would now require a larger portion due to degraded channel quality. The product of $r(f,A) \times \delta(f,C)$, with $\delta(f,C) < 1$, captures this effect in terms of the effective amount of resource units consumed by $f$ at $C$. As $d_{AC}$ increases, $throughput(d_{AC})$ decreases, and $\delta(f,C)$ decreases. 
We show in Appendix\ref{sec:app-analysis} how the specific value of $\delta(f,C)$ in our formulation closely approximates the amount of resource units consumed at $C$ in many cases (and slightly under- or over-estimates it in others). We also validate our formulation using real-world experiments (\S\ref{sec:validation}), showing that it achieves good enough approximation. 


\paragraphb{Interference from multiple sources} A device $C$ may see interference from multiple flows $f_1,....f_k$. We approximate the effect of interference from multiple such sources by simply summing up the amount of resource units consumed at $C$ individually by each such interfering flow. This provides a very close approximation, especially in cases where the interference sources are closeby and the effects of interference are higher. It may slightly over-estimate the effects of interference in other cases -- we provide detailed analysis in Appendix~\ref{sec:app-analysis}. 


\paragraphb{Multi-hop paths} We use the above formulations to also reason about the resource units consumed by flow $f$ on multi-hop paths. Transmission over a given hop will directly consume resources at the source and destination device for that hop, and will additionally consume resources at other hops in the flow's path due to interference. 


\begin{figure}
\centering
\captionsetup{justification=centering}
\captionsetup[subfigure]{justification=centering}
\centering
\subfloat[Actual vs Predicted Throughput for Flow A]{\includegraphics[width=0.30\textwidth]{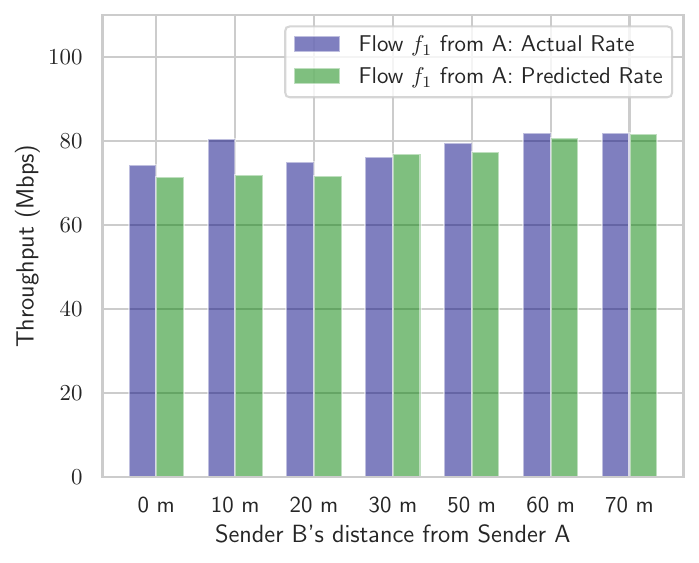} \label{fig:interference-exp-res}}
\subfloat[Experiment Setup]{\includegraphics[width=0.16\textwidth]{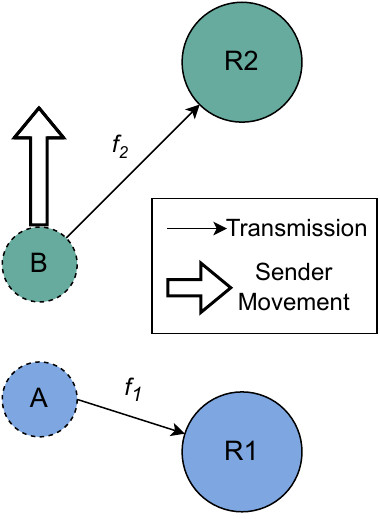} \label{fig:interference-exp-setup}}
\vspace{-10pt}
\caption{Validating our predicted interference effects from Sender B's transmission to R2 on Sender A's flow to R1 }
\vspace{-10pt}
\label{fig:interference-abstraction}
\end{figure}

\begin{figure}
\centering
\captionsetup{justification=centering}
\captionsetup[subfigure]{justification=centering}
\centering
\subfloat[Resource units from both flows add up to 1]{\includegraphics[width=0.23\textwidth]{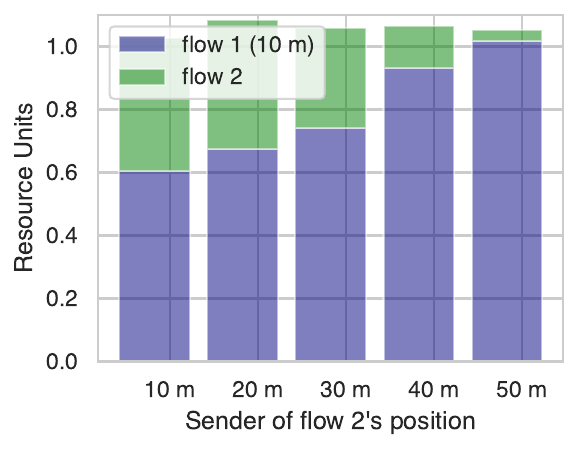} \label{fig:distance-throughput-unfair}}
\subfloat[We can divide resource units in any fraction between flows (1:1 here)]{\includegraphics[width=0.23\textwidth]{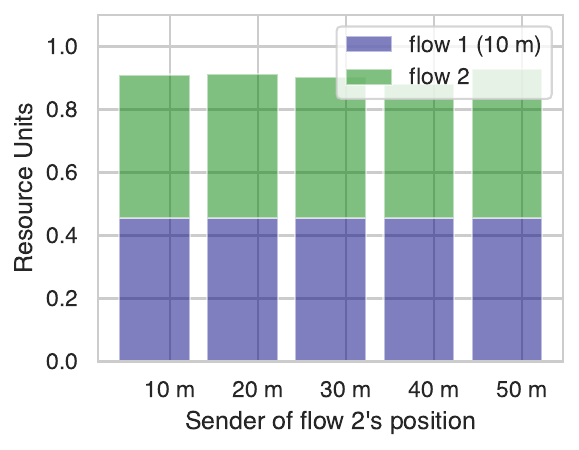} \label{fig:distance-throughput-fair}}
\vspace{-10pt}
\caption{Micro benchmark highlights we need explicit rate enforcement.}
\vspace{-10pt}
\label{fig:distance-abstraction}
\end{figure}

\subsection{Experimental Validation and Insights} 
\label{sec:validation}

We validate the correctness of our resource unit formulation using a series of real-world experiments in the cornfield.

\paragraphb{Validating interference computation} 
We place a sender $A$ at distance of $d_{AR_1} = 10$ m from router $R_1$. We place another router $R_2$ at distance of 10 m from $R_1$. We then vary the position of sender $B$ that transmits to $R_2$, increasing its distance $d_{AB}$ from $A$ across different experiment runs, as shown in Figure \ref{fig:interference-exp-setup}. Doing so also varies the distance $d_{BR_2}$. We fix the datarate of flow $f_2$ from $B$ to $R_2$ and use our formulation to compute the amount of resource units consumed by $f_2$ on $A$ and $R_1$ due to interference. We the compute the corresponding datarate achieved by flow $f_1$ from $A$ to $R_1$ that uses the leftover resource units. Figure \ref{fig:interference-exp-res} shows how well our predicted throughput matches the observed throughput of $f_1$.~\footnote{We assigned a smaller than fair share rate to $f_2$ (with the values varying with distance) and validated that it was able to achieve it, inspite of interference from $f_1$.}

\paragraphb{Need for explicit rate enforcement}
In another experiment, we set up two devices $A$ and $B$ that simultaneously transmit data to a router $R$. We keep the distance between $A$ and $R$ fixed, and vary the distance between $B$ and $R$, increasing it across different runs. We measure the datarate achieved by each flow, and compute the corresponding resource units consumed by each flow using Equation~\ref{eqn:throughput-distance}. Figure \ref{fig:distance-throughput-unfair} shows how the resource units consumed by each flow indeed add up to one (which validates our basic abstraction of resource units) -- the channel bandwidth available at $R$ is split between the two flows. We further observe that the datarate naturally achieved by the two flows neither corresponds to fair use of channel capacity nor to fairness in terms of datarates (the transmission from $B$ consumes increasingly smaller amount of resource units as the distance between $B$ and $R$ increases). 

Guided by this observation, our TE systems makes flow rate enforcement a first-class citizen. It uses our formulation in \S\ref{sec:resource-unit} to reason about the amount of resource units that transmissions at a given datarates and at different distances must consume, and then explicitly computes flow rates that adhere to the specified policy (e.g. fair division of channel bandwidth between contending flows, prioritizing one flow over another, etc). Figure \ref{fig:distance-throughput-fair} shows that when we apply such reasoning to fix the datarates of the transmissions (with 10\% headroom), we can realize the desired policy (fair division of channel capacity in this case). Such explicit per-flow rate enforcement thereby allows us to control how the wireless network capacity is shared across flows in a manner that is agnostic of the underlying mechanisms (TDMA, OFDMA, etc), and enables us to cleanly reason about the resource units consumed by different flow allocation decisions.





\subsection{Traffic Engineering Algorithm}
\label{sec:heuristics}


We can use our resource unit abstraction to formulate our TE decisions as a multi-commodity flow problem (MCFP) \cite{even1975mcf} (similar to how WAN TE is modeled~\cite{hong2013achieving}). 
However, unlike WAN traffic engineering, that operates on flow aggregates that can be split across multiple paths as per the output of the MCFP, \farmernetes operates at the scale of individual flows that must use a single path over the WiFi mesh to avoid extensive re-ordering. MCFP is known to be NP-hard if flow splitting is not allowed. Moreover, our MCFP extension for incorporating channel allocation reduces to a graph-coloring problem which is also NP-hard. We therefore develop a series of greedy algorithms to configure our TE knobs.

At each invocation, \farmernetes TE engine first picks real-time tasks that must be scheduled in that invocation and assigns them network resources in both the 2.4GHz and 5GHz tier (as described below). It then configures routes and rates for the non real-time (data collection) tasks, only assigning them network resources spared by the real-time flows.  

\paragraphb{Picking real-time flows to schedule} The TE engine computes the \emph{slack} time for each real-time flow by subtracting its remaining duration from the deadline of the task. It first picks the flows with zero slack for scheduling, greedily allocating them network resources in both the 2.4GHz and 5GHz tiers (as detailed later in this section). For each of the remaining real-time flows with non-zero slack (ordered by their slack time from lowest to highest), the TE engine iteratively checks to see if the flow can be scheduled by greedily allocating it resources in the 2.4GHz and 5GHz tiers (i.e. if the resources allocated to it across both tiers, using the techniques described below, are sufficient to meet its demand). If yes, the TE engine schedules the flow and accounts for its network resource consumption in the two tiers, before moving over to do the same check for the next flow in the ordered list. 


\paragraphb{Resource allocation in 2.4GHz tier} \emph{(Access point selection and channel configuration):} As mentioned above, the TE engine greedily allocates resources in the 2.4GHz tier, one  flow at a time. 
For a given flow $f$, the TE engine iterates over all possible APs in the range of its source, and picks the one that minimizes the function $F(AP)$ specified below. If there is already a flow scheduled on an AP, it implies that the channel for that AP has been pre-allocated, and the same channel is used for the flow when computing the value of the function. When iterating over APs that have not yet been assigned a flow (and a channel), the TE engine additionally iterates over the three possible channel in 2.4GHz band, and picks the channel that minimizes $F(AP)$. We define $F(AP)$ as follows:

    $$F(AP) = r(f, AP) + C(AP) \\ + \sum_{R_j \in Neighbors(S)} (r(f, R_j) + C(R_j))$$
Here, $r(f, AP)$ is the amount of resource units the real-time flow $f$ (sending at its desired rate from its source $S$) would consume at access point $AP$ (if it is selected). The second term 
$C(AP) = \sum_{f_i \neq f}r(f_i, AP)$ is existing contention at AP and it captures the amount of resource units consumed at $AP$ by other flows that have already been scheduled. $Neighbours(S)$ captures the set of devices that are in the range of the sender and share the same channel. The final term therefore captures the amount of resource units consumed at these routers by flows that have already been scheduled ($C(R_j)$), and the additional amount of resource units that $f$ would consume on them due to interference ($r(f,R_j)$). This term primarily guides channel selection. Thus, the choice of AP and the corresponding channel are dictated by a combination of the AP's distance from the source (that would minimize $r(f, AP)$), how contended the AP is, and how contended the neighboring APs on the same channel (that would suffer interference) are.~\footnote{When multiple flows with zero slack are to be scheduled together, \farmernetes' TE engine sorts them by their demand (or desired rates) before greedily allocating the 2.4GHz resources. This helps with improving efficiency, since a flow with higher demand will first get assigned to an AP that is nearer, thereby enabling it to use lesser amount of resource units than what it would have used with a more distant AP. As mentioned earlier, flows with non-zero slack are assigned resources in the order of their remaining slack. 
}

\paragraphb{Resource allocation in 5GHz tier} \emph{(Route Computation):} After selecting the AP and the channel at the 2.4GHz tier, the TE engine computes the flow's route on the 5GHz tier. At each router hop in the grid topology, \farmernetes restricts the choice of the next hop to the nearest four routers. In other words, the flow is restricted to picking a Manhattan route.~\footnote{We found that including more distant routers in the search space for the next hop or routing along the diagonals of the grids resulted in lower network throughput due to the combined effect of more amount of resource units consumed by routing along more distant hops and the interference of such transmissions on neighboring and intermediate routers.} The TE engine computes a weight for each router in the grid using the following function: 
$$ w(R) = \max(0, C(R) + 2r(f,R) -1)  + $$ 
$$ \sum_{R_n \in Neighbors(R)} \max(0, C(R_n) + r(f, R_n) - 1) $$



The TE engine then picks the shortest weighted Manhattan route for the flow. 
The router weight $w(R)$ captures existing contention on router $R$ and its neighbors and the additional interference that routing flow $f$ through $R$ will cause. We further account for the fact that a router $R$ will both receive and send the flow, consuming twice $r(f,R)$ amount of resource units.~\footnote{We can compute $r(f,R)$ without first determining the route, since $f$ will be transmitted to $R$ from (and by $R$ to) one of its four equidistant neighbors in the grid.} The weight function favors a shorter path that has spare capacity (at both router $R$ and its neighbors)  over a longer less contended path. The second term in the two $max$ functions captures the over-commitment in resource units (i.e. these terms are less than zero when there is spare capacity). 

\paragraphb{Rate enforcement for real-time flows} 
A real-time flow gets its desired rate when sufficient amount of resource units are available. However, situations can arise where a real-time flow with zero slack must be scheduled on a contended network, that does not have enough resources to accommodate its desired datarate. In such cases, the available capacity in terms of throughput (Mbps) is divided fairly between the flows, and flows are configured to send at the specified rate.

\paragraphb{Rate Enforcement for Non-real-time flows} Once the real-time flows have been allocated network resources, we allocate the spare network resources to the non-real time (data collection) tasks using a simple water-filling algorithm. 
We randomly pick a flow from the list of non-real-time flows, and select the nearest AP and shortest path by $w(R)$ cost function for it. We compute the datarate that can be assigned to this flow, such that it can consume the left-over resource units on the bottleneck device along its path and in its interference range.
We then proceed to randomly select the next non-real-time flow, and similarly assign it a path and sending rate whenever there are left-over resources. 
Randomness in the order in which these flows are picked ensures fair scheduling across these flows (that have lax deadlines) in the longer run. 
Our TE engine also supports other policies, e.g. prioritizing the non-realtime flows differently based on the remaining amount of data they must send and their deadlines.

\paragraphb{Handling Mobility}
Many real-time tasks in a farm setting would be executed by static devices (e.g. a robot de-weeding a crop or picking berries at a static location). \farmernetes also deals with device mobility, by leveraging the fact that robots in a farm setting move along simple trajectories (between crop rows) and at low speed (a few meters per minute). To handle changes in resource allocation decisions as a device moves,  the TE engine computes the TE knobs in each invocation over a series of epochs, based on the expected trajectory of the devices. For example, it computes that device $X$ should switch its connection from router $A$ to router $B$ after 10 epochs, router $B$ should switch to channel $c$ after 9 epochs~\footnote{We observed that a device becomes unavailable for 5s when switching channel (Appendix \ref{sec:channel-switching}), so our TE engine strives to avoid channel switching on a mobile device while it is transmitting a real-time flow, and instead re-configures the router to which the device is handed over to use same channel as the device.}, or that the device sending a non-real-time flow to $B$ should reduce its sending rate after 9 epochs, etc. Potential errors in such pre-computed decisions, caused by deviations in the expected trajectory, can be corrected in the subsequent invocation of the TE engine.

\paragraphb{Deviations in expected behavior} 
Our TE decisions rely on data-driven estimates of on-farm throughput at varying distances (\S\ref{sec:resource-unit}). The actual throughput can deviate from these estimates due to various factors (weather on a particular day, damaged equipments, etc). We can extend \farmernetes' TE engine to infer such deviations from observed performance, and correct the TE decisions in subsequent invocations. We can also add greater robustness to such deviations by leaving some headroom when assigning rates to the non real time flows. Small deviations from estimated performance also arise due to spatial variations (as shown in \S\ref{sec:measurements}) --  our evaluation in \S\ref{sec:eval} shows the overall effectiveness of \farmernetes' TE decisions inspite of these spatial variations.  
\section{Evaluation}
\label{sec:eval}

\begin{figure*}
\centering
\captionsetup{justification=centering}
\captionsetup[subfigure]{justification=centering}
\centering
\subfloat[Overall Throughput]{\includegraphics[width=0.245\textwidth]{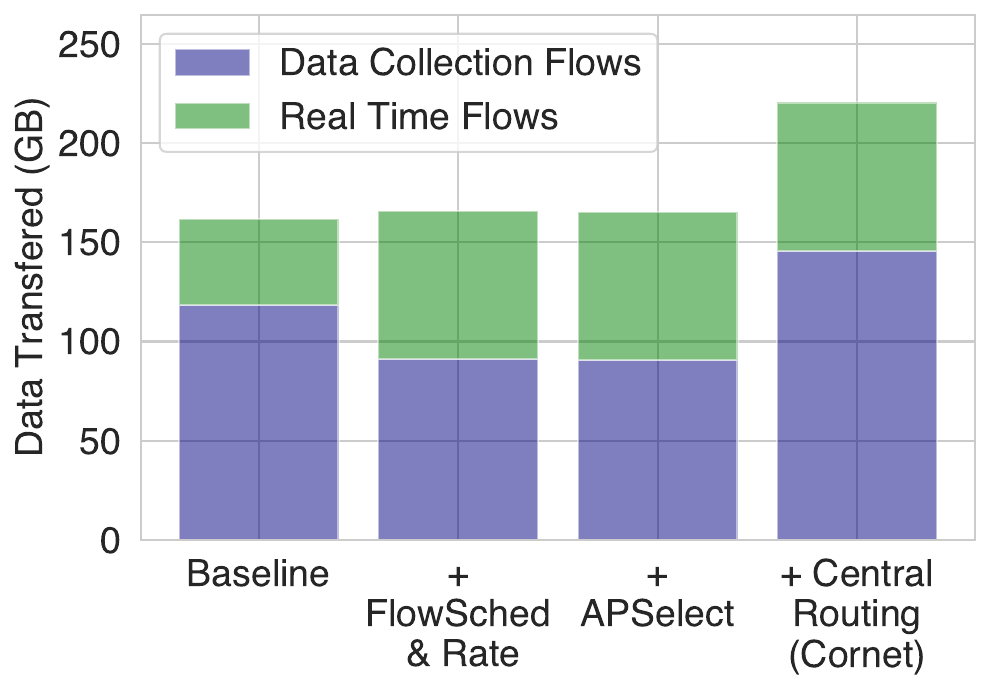} \label{fig:non_realtime_data}}
\subfloat[Per-flow Throughput for Real-time flows]{\includegraphics[width=0.245\textwidth]{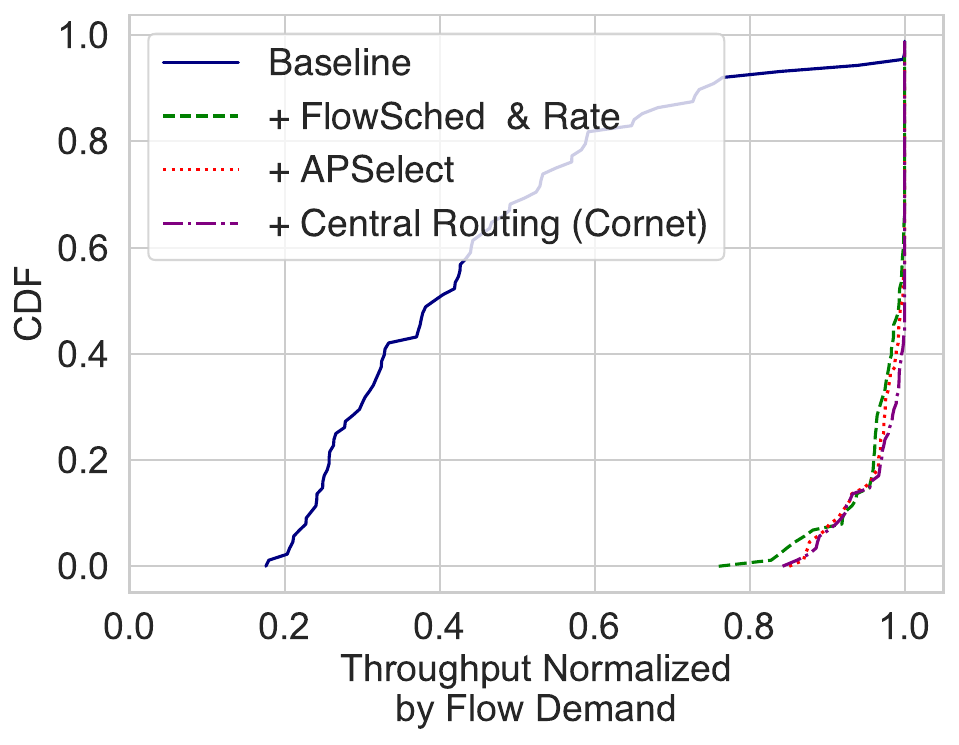} \label{fig:realtime_qos}}
\subfloat[Overall Throughput]{\includegraphics[width=0.245\textwidth]{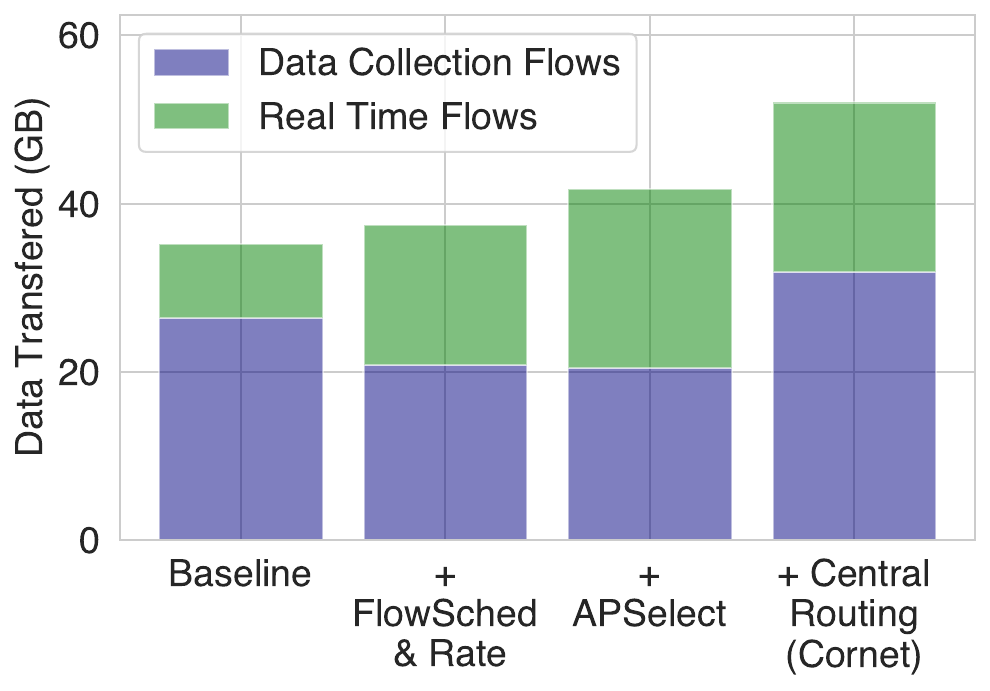} \label{fig:local_data}}
\subfloat[Per-flow Throughput for Real-time flows]{\includegraphics[width=0.245\textwidth]{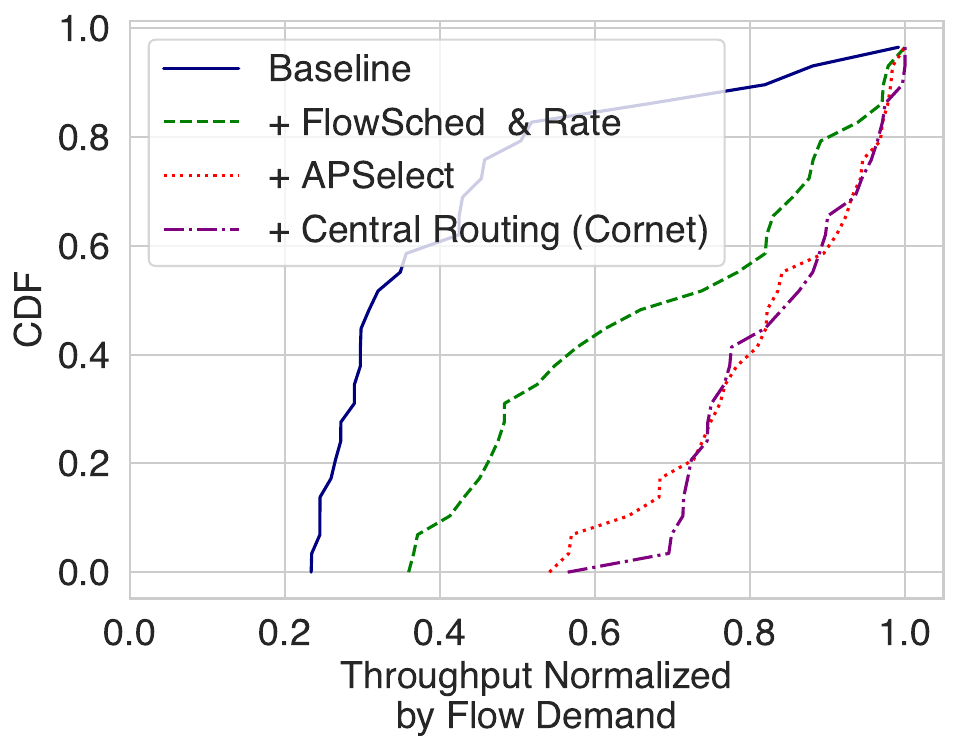} \label{fig:local_qos}}
\vspace{-10pt}
\caption{Improvements in throughput and per-flow QoS as we incrementally add \farmernetes' algorithms to the baseline. (a) and (b) is for a large scale deployment,  (c) and (d) for a more localized workload}
\vspace{-10pt}
\label{fig:eval_1}
\end{figure*}

We use large-scale trace-driven simulations (\S\ref{sec:trace-eval}) and small-scale real-world experiments (\S\ref{sec:real_world}) to evaluate \farmernetes'. 

\subsection{Trace-driven Simulations} 
\label{sec:trace-eval}

\paragraphb{Simulator} We build a simulator to simulate \farmernetes' two-tiered mesh network. It can be configured with an input workload, comprising of source devices, i.e. sensors and receivers, at specified locations, trajectories for mobile devices and the corresponding flow demands for both real-time and data collection tasks. It implements \farmernetes TE techniques at both tiers, along with other baselines that we compare against. Our simulation uses our measurement traces and the formulations described in \S\ref{sec:resource-unit} to compute flow throughput (including the effects of interference).~\footnote{Our simulator models ideal spectrum sharing, by using max-min fairness to split resource units at a device among contending flows.} We additionally model spatial variations in throughput by scaling the computed throughput with a randomly generated multiplier that produces the same standard deviation as that observed in \S\ref{sec:measurements}.

\paragraphb{Baselines} We compare \farmernetes against a naive mesh baseline that schedules flows as and when they are requested, picks the nearest AP at the 2.4GHz over a randomly selected channel, and routes along the shortest Manhattan path in the 5GHz tier (picking a path at random among multiple choices of the same path length in order to appropriately load-balance the traffic). A wireless mesh routing designed to pick better quality links (as in Roofnet~\cite{aguayo2004roofnet, etx-roofnet}) over shortest path by hop count, would pick a similar set of paths as our naive mesh baseline. 

We incrementally augment this baseline with different elements of \farmernetes' TE to evaluate their impact: (i) In \emph{+ FlowSched \& RateAssign}, we schedule real-time flows based on their slack time, but still select the nearest AP at 2.4GHz tier and use the naive shortest-path Manhattan routing in 5GHz tier. We assign rates to the non real-time flows so as to only use the spare capacity. (ii) \emph{+ APSelect} augments the above by using \farmernetes' AP and channel selection logic at the 2.4GHz tier instead of picking the nearest AP (iii) \emph{+ CentralRouting}, further augments +APSelect by replacing  the shortest-path Manhattan routing with \farmernetes centralized contention-aware routing, thus incorporating all elements of \farmernetes' TE.


\paragraphb{Scenario 1: Large scale Deployment}
We simulate a square-shaped 450 acre that has 225 routers placed in a 15 $\times$ 15 grid where each router is 90 meters apart. Three routers in the first-row act as the gateway routers to the edge and are the destination for all flows discussed ahead.
We simulate a duration of about 4 hours (250 minutes). Based on \S\ref{sec:workload}, our input farm workload over the simulated duration contains a total of: (i) 40 daily data collection tasks (e.g. for thermal and RGB data), each of which require collecting 15MB data in the first 200 minutes. We also include 8 weekly data collection tasks (e.g. for hyperspectral images) each of which require collecting 500 MB in the next two weeks. (ii) 100 real-time tasks, each with a demands between 10 to 20 Mbps, duration between 5 and 20 minutes. The deadline is such that they can be scheduled 0-10 minutes  after they have been requests. These task requests are distributed randomly over space. About 30\% of the tasks originate from mobile devices, where a mobile device travels a distance of 5-10 meters per minute along the crop row over the duration of the task. At any time, at most 15 real-time flows are active.
We assumes all tasks are pre-emptible, i.e. they can be paused and resumed, as long as they finish by their deadline.

\paragraphi{Results:}
Figure \ref{fig:non_realtime_data} and \ref{fig:realtime_qos} shows the impact of different element of \farmernetes. Figure \ref{fig:non_realtime_data} shows the total amount of data transferred across the two types of tasks. Figure \ref{fig:realtime_qos} shows the cumulative distribution function (CDF) of the throughput achieved by each real-time flow, normalized by its demand. \farmernetes achieves better network utilization than the baseline (transferring 1.4 $\times$ more amount of data), and it is able to better prioritize more urgent real-time tasks (resulting in 2.5 $\times$ higher normalized throughput for the real-time flows on average, in most cases satisfying flow demands completely). Smarter flow scheduling and rate assignment enabled better performance for real-time flows (allowing them to better meet their target throughput). The relative impact of AP selection and channel allocation at the 2.4GHz tier was small in this setting, since simultaneous real-time flows were sourced at devices spread across the entire farm. \farmernetes' contention-aware routing selects better paths in the 5GHz tier, increasing overall network utilization.

\begin{figure}
\centering
\captionsetup{justification=centering}
\captionsetup[subfigure]{justification=centering}
\centering
\subfloat[Overall Throughput]{\includegraphics[width=0.23\textwidth]{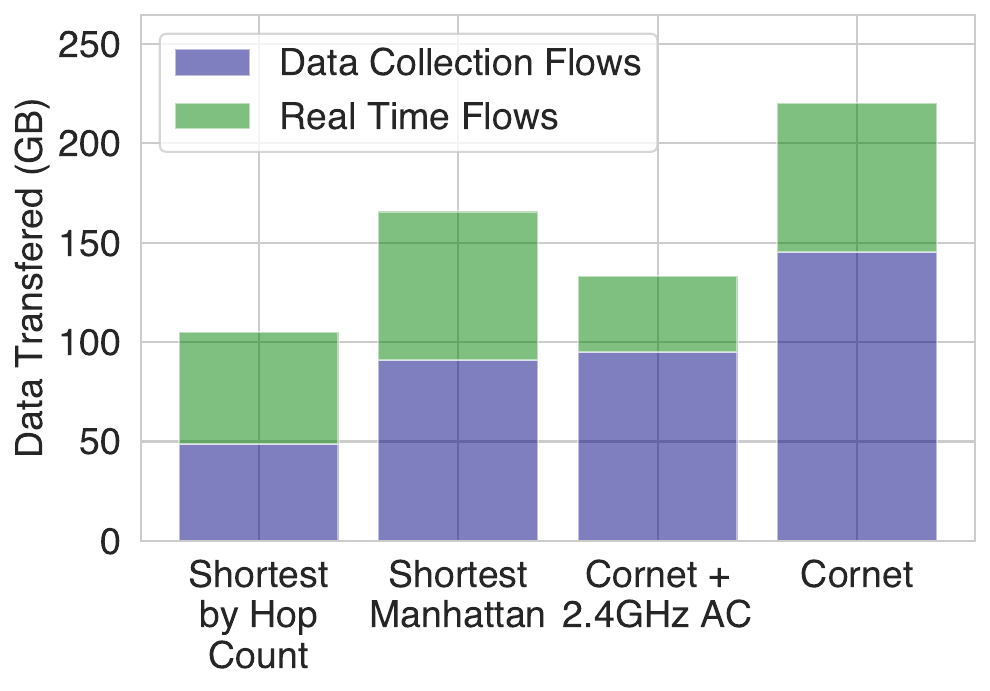} \label{fig:paths-data}}
\subfloat[Per-flow Throughput for Real-time flows]{\includegraphics[width=0.23\textwidth]{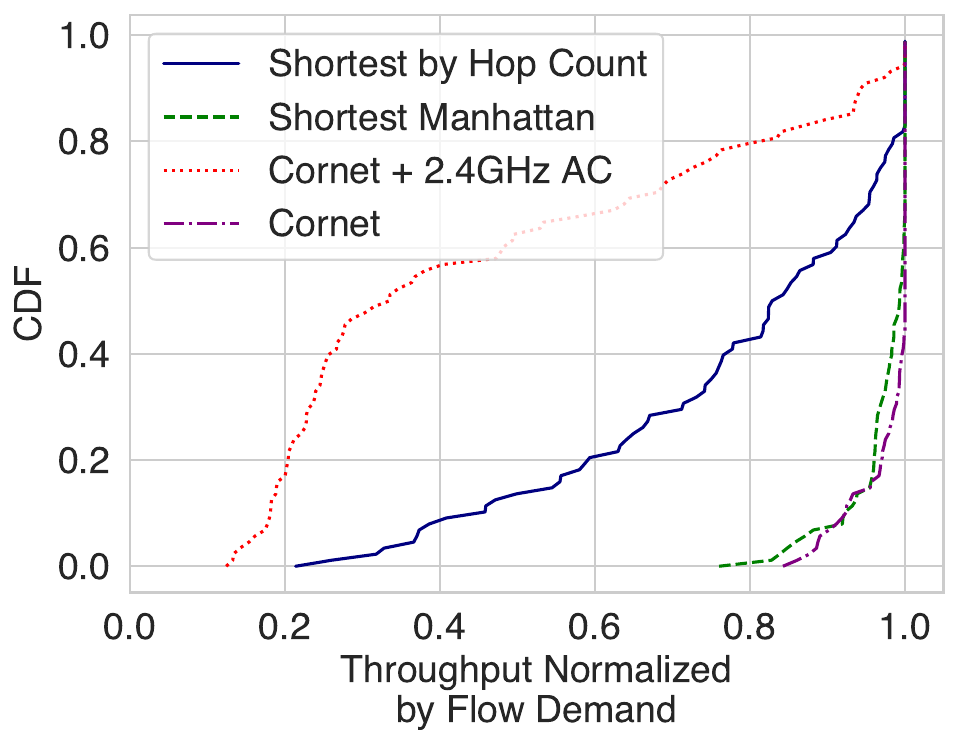} \label{fig:paths-qos}}
\vspace{-10pt}
\caption{\farmernetes' routing algorithm compared with some other baselines.}
\vspace{-10pt}
\label{fig:paths-algos}
\end{figure}

\paragraphb{Scenario 2: Localized Operations} 
 We next simulated a scenario where most real-time tasks originate from a certain area e.g. multiple robots picking berries in a specific area. In this workload, alongside non-real-time flows, we have multiple sets of up to 5 robots in proximity performing real-time tasks. We find higher relative benefits of AP selection and channel allocation in this case (figures \ref{fig:local_data} \& \ref{fig:local_qos}).
 


\paragraphb{Comparison with other routing mechanisms}
We compare \farmernetes' contention-aware routing with: (i) Shortest path by hop count (a popular strategy for multi-hop wireless networks~\cite{chakeres2004aodv, johnson2001dsr, dsdv, olsr}).  (ii) Shortest Manhattan path (the same routing strategy used as a baseline in our experiments above) favors better quality links over simply minimizing hop count (as in~\cite{aguayo2004roofnet, etx-roofnet}). (iii) \farmernetes when allowed to use 2.4GHz band above canopy to augment 5GHz (which, for reasons discussed in \S\ref{sec:two-tier}, ends up adversely effecting performance). We find that \farmernetes' strictly tiered contention-aware routing on 5GHz outperforms each of these baselines (Figure~\ref{fig:paths-algos}).

\begin{figure}
\centering
\captionsetup{justification=centering}
\captionsetup[subfigure]{justification=centering}
\centering
\includegraphics[width=0.3\textwidth]{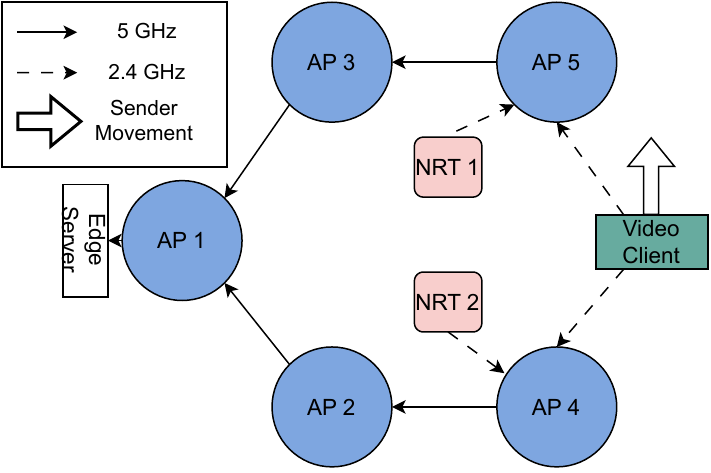} 
\vspace{-10pt}
\caption{Setup for our real world deployment of \farmernetes}
\vspace{-10pt}
\label{fig:real-world-setup}
\end{figure}

\subsection{Real World Evaluation}
\label{sec:real_world}

\paragraphb{Setup} We also evaluated the effectiveness of \farmernetes TE decisions using a small-scale real-world deployment. We used five ASUS RT-AC86U routers (updated with Asuswrt Merlin firmware which allowed us to remotely reconfigure their routes and channels), that were mounted on poles and placed at different locations 
on the farm, and created a mesh network using 5 GHz band (see figure \ref{fig:real-world-setup}). One router was connected to an edge server using Ethernet. We emulated non-real-time (data collection) flows on two under-canopy devices that sent data to the edge server using iperf. These devices connected to the specified routers over 2.4 GHz.
Another device sent a video stream to the server, where the video required a bitrate of 15 Mbps. We measured the frame drop rate for the video at the edge server for each run. The positions of non-real-time flow sources were fixed, whereas the video source's location was varied. 

\begin{figure}
\centering
\captionsetup{justification=centering}
\captionsetup[subfigure]{justification=centering}
\centering
\subfloat[Naive Wifi Mesh]{\includegraphics[width=0.23\textwidth]{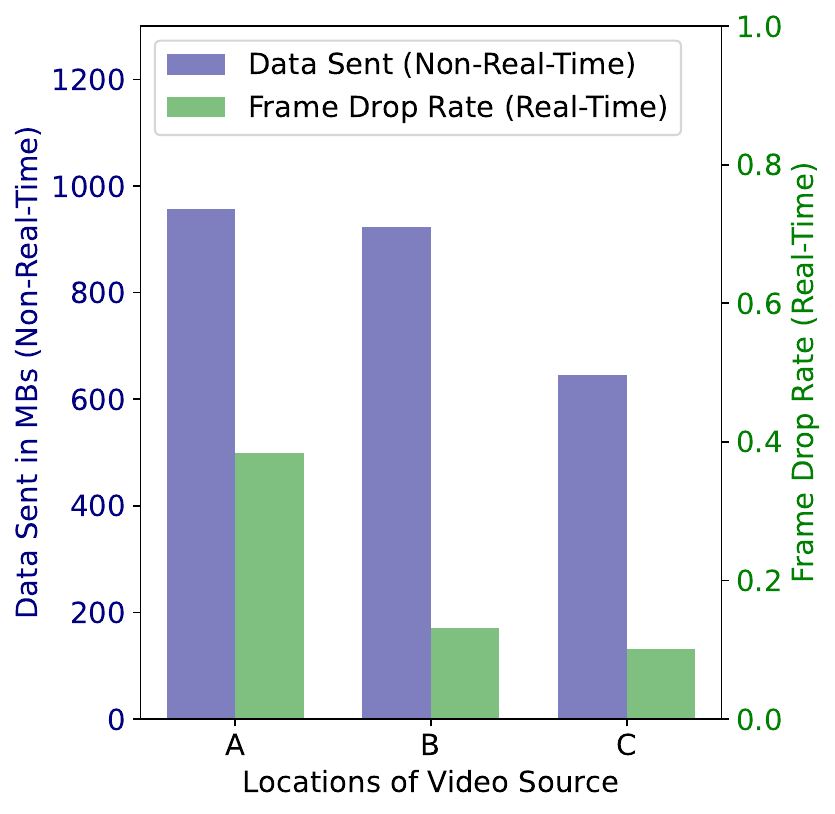} \label{fig:real-world-naive}}
\subfloat[\farmernetes]{\includegraphics[width=0.23\textwidth]{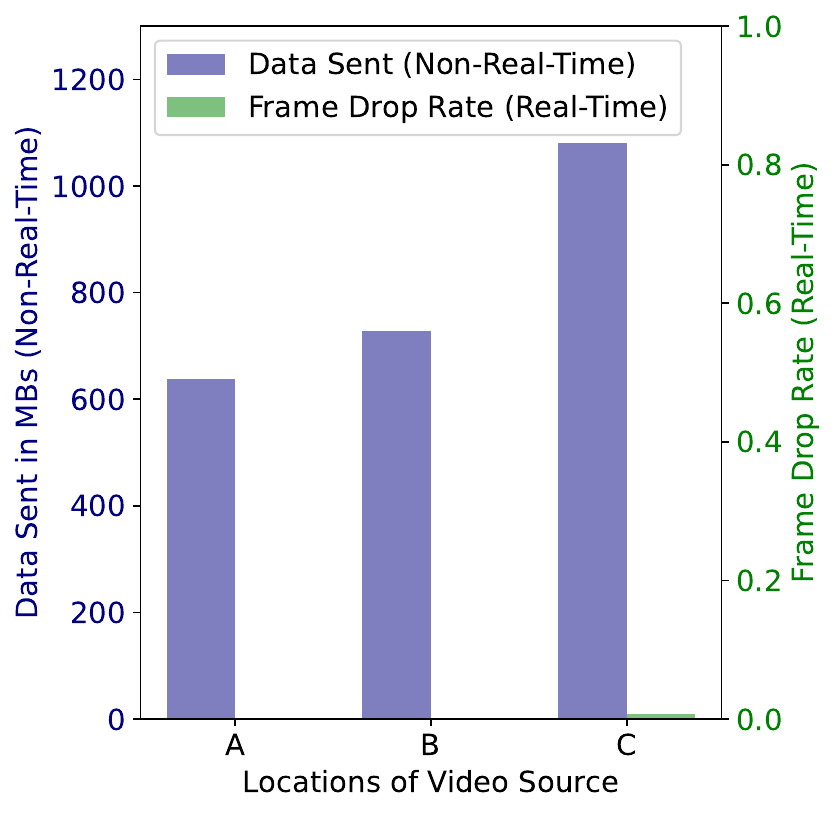} \label{fig:real-world-farmernetes}}
\vspace{-10pt}
\caption{\farmernetes results in improved network management in real-world deployment.}
\vspace{-15pt}
\label{fig:realworld-deploy}
\end{figure}

\paragraphb{Results} Figure \ref{fig:realworld-deploy} shows the performance at different locations of the video flow, comparing \farmernetes with the naive WiFi mesh baseline. Given the relatively small number of routers in our setup, \farmernetes picked the same (optimal) paths for each flow as the naive baseline. However, \farmernetes made smarter decisions about channel allocation, AP selection, and especially rate assignment. We observe how, with the naive baseline, video flow suffers from high drop rates as non-real-time flows hog all the bandwidth. \farmernetes ensures almost zero drop rate and still ensures high enough throughput for non-real-time flows. In some cases (e.g. when the video source was at location $C$), the overall throughput was better with \farmernetes because of careful channel allocation. 

\section{Related Work}


\paragraphb{Farm Networking} FarmBeats~\cite{vasisht2017farmbeats} proposes an architecture where TVWS base-stations (located at the edges of the field) provide long range connectivity to the Internet gateway at the farmer's office. On-field sensors connect to the nearest TVWS base-stations over WiFi. Our measurements reveal the difficulty in scaling such an architecture, with the WiFi range for supporting sufficiently high bandwidth being limited to a few tens of meters (particular under canopy). \farmernetes' multi-hop mesh network and traffic management can be combined with the FarmBeats architecture for enabling better connectivity between on-field devices and the TVWS base-stations. TVWS-Whipser~\cite{chakraborty2022whisper} enables long-range connectivity for IoT devices over TVWS, but similar to LoRAWAN, can only support throughput of few 10s of Kbps.  

Wu et. al~\cite{zigbee-undercanopy} measure ZigBee performance (receive signal strength and packet loss rate) at 2.4GHz and 433MHz in a farm setting at varying transmission heights above the crop canopy. Brinkhoff et. al. \cite{brinkhoff2017characterization} measure the range and signal strength of WiFi 2.4 GHz in a rice field for transmissions at varying height above the crop canopy. Driven by the requirements of emerging farming techniques that require under-canopy operation, we conduct the first-of-its-kind under-canopy measurement measurement of WiFi throughput,  comparing across both 2.4GHz and 5GHZ bands, and use the results to guide our farm network design.

\paragraphb{Wireless Mesh} Routing and traffic management for multi-hop wireless mesh networks, designed for settings with unpredictable traffic patterns and external interference, have typically used decentralized protocols to compute routes with shortest paths~\cite{chakeres2004aodv, johnson2001dsr, dsdv, olsr} or better link qualities~\cite{aguayo2004roofnet, etx-roofnet}. Similarly, existing channel assignment algorithms are also reactive because of unpredictable traffic patterns of typical WiFi networks \cite{rozner2007traffic}.
There have been a few proposals that formulate centralized routing problems for WiFi mesh with the objective of energy conservation~\cite{amokrane} and channel assignment for multi-band routers~\cite{raniwala2004centralized}. These proposals side-step the key challenges in reasoning about wireless link capacity by making several simplifying assumptions, e.g. modeling wireless capacity in terms of absolute datarate (in bits/s), ignoring the impact of transmission distance, and modeling interference as a binary effect (whether or not two links interfere) agnostic of the distance between them. To the best of our knowledge, ours is the first centralized traffic engineering system for wireless mesh that systematically accounts for such effects and also validates them using real-world experiments. 

There is ongoing effort towards enabling wireless time-sensitive networking (TSN)~\cite{wtsn, wtsn2}, that involve using a centralized controller for better time synchronization between transmitting devices and frame-by-frame scheduling. These layer-2 efforts, yet to be realized in practice, target the ambitious goal of achieving higher reliability and predictability in how the underlying wireless spectrum gets shared by contending flows, and are orthogonal and complementary to \farmernetes' flow-level traffic engineering decisions (that are agnostic of the underlying spectrum sharing mechanisms through explicit rate enforcement). 

There is an extensive work on modeling the effects of interference \cite{qiu2007general, garetto2005modeling, gupta2000capacity, jain2003impact}, that model a flow's throughput based on the channel quality (which depends on transmission distance and interference). While we draw insights from these work, we flip the conventional reasoning and ask how much wireless network resources a flow sending at a given datarate along a given path would consume on different devices, and show how such reasoning can enable TE decisions. 




\section{Discussion and Future Work}
\label{sec:discussion}



\paragraphb{Above-canopy sensors} We largely focused on using the 5GHz above-canopy tier in \farmernetes for routing the data generated by under-canopy sensors and robots. The 5GHz tier can additionally be used to support above-canopy sensors (e.g. cameras on poles and drones). Some amount of bandwidth in the 5GHz tier can be reserved to support high-speed mobility of tele-operated drones whenever such tasks are requested.

\paragraphb{Power availability} We assume on-field routers and devices are solar powered with sufficient battery capacity.
Under limited battery life, routers can be duty-cycled to save power (thereby reducing the effective network capacity)~\cite{vasisht2017farmbeats}. \farmernetes' TE can be extended to incorporate such factors.




\paragraphb{Deployment challenges} 
Deploying and maintaining a WiFi mesh in farms can be challenging in practice. Similar challenges would hold for deploying smart agricultural devices more generally. Despite these challenges, we are already seeing deployments of smart agriculture and networking infrastructure on farms (e.g.~\cite{farmproven-tech}). We can also envision rise of dedicated farming infrastructure services (e.g.~\cite{ayrstone_productivity_2023, earthsense}) that provide farmers with the required equipments, along with assistance in deploying and maintaining them. Nonetheless, easing deployment and enabling automated maintainence of on-farm network infrastructure remains an interesting open challenge. 



\section{Conclusion}

In this work, we examine the scope of using WiFi for meeting the high bandwidth requirements of emerging smart farming techniques. We use insights from extensive WiFi measurements in a cornfield to design \farmernetes, a WiFi mesh network for farms. 
Our quest for improving on-farm networking leads to a new way of reasoning about wireless network capacity that enables centralized traffic engineering decisions. Going forward, it will be interesting to see how our design extends to other domains that use a wireless network to support schedulable workloads (e.g. mining, factories, etc).    


\bibliographystyle{plain}
\bibliography{reference}

\appendix

\section{Extended Analysis on Computing Consumed Resource Units}
\label{sec:app-analysis}

We refer to figure in \S\ref{sec:resource-unit} in the main paper to provide some more detailed analysis of our formulation for computing the amount of resource units consumed by a flow due to interference. 

\paragraphb{Analysis for a single source of interference.} We focus on computing the resource units consumed by flow $f$ from $A$ to $B$ with datarate XMbps at $C$. 
We discussed the case where $d_{AC} \leq d_{AB}$ in \S\ref{sec:resource-unit}. We now discuss the case where $d_{AC} > d_{AB}$ and the impact of interference is realized through reduced signal-to-noise ratio. 

 \begin{subequations}
 \begin{align}
    r(f, C) =& r(f,A) \times \delta(f,C) \\
    =& r(f,A) \times \frac{throughput(d_{AC})}{throughput(d_{AB})} \\
    =& r(f,A) \times \frac{log(ss(d_{AC}))}{log(ss(d_{AB})} \label{equ:shannon1}
\end{align}
\end{subequations}

We get Equation~\ref{equ:shannon1} by applying Shannon's Law under the simplifying assumption that signal strength is strong enough to ignore noise floor (i.e.  $throughput(d) \approx BWlog(ss(d)$). Here, $BW$ is the total channel bandwidth available for the transmission, and $ss(d)$ is the signal strength at distance $d$ (in the absence of any interference). 

The specific value of $\delta(f,C)$ in our formulation closely approximates the amount of resource units consumed at $C$ when we reason about the interference effects from $f$ in terms of using the leftover resource units at $C$ for another flow $f'$ from $C$ to $D$ at distance $d_{CD} = d_{AB}$
As per our formulation, the datarate achieved by $f'$ will be its share of channel bandwidth multiplied by its signal strength in dB ($log(ss(d_{CD}))$). This will be equal to:
\begin{subequations}
\begin{align}
&= (1 - r(f,C))BWlog(ss(d_{CD})) \\
&= \left( 1 - r(f,A) \times \frac{log(ss(d_{AC}))}{log(ss(d_{AB})} \right) BWlog(ss(d_{CD}) \\
&= (1 - r(f,A)) BWlog(ss(d_{CD}) \\
&+ r(f,A) BWlog(ss(d_{CD}) \left( 1 - \frac{log(ss(d_{AC}))}{log(ss(d_{AB})} \right)
\end{align}
\end{subequations}

When $d_{CD}$ = $d_{AB}$, this is equal to: 
\begin{subequations}
\begin{align}
&= (1 - r(f,A)) BWlog(ss(d_{CD}) \\
&+ r(f,A) (BWlog(ss(d_{CD}) - BWlog(log(ss(d_{AC})) \\
&= (1 - r(f,A)) BWlog(ss(d_{CD}) \\
&+ r(f,A) BWlog \left( \frac{ss(d_{CD})}{ss(d_{AC})} \right)
\end{align}
\end{subequations}
This correctly captures the degradation in SINR (signal to interference-noise ratio) for the transmission from $C$ to $D$ due to interference power of strength $ss(d_{AC})$ on the portion of channel bandwidth ($r(f,A)$) accessed by the interfering flow $f$. 
For $d_{CD} > d_{AB}$ and $d_{CD} < d_{AB}$, our formulation ends up overestimating and under-estimating $r(f,C)$ respectively.

\paragraphb{Analysis for multiple sources of interference.}
A device $C$ may see interference from multiple flows $f_1,....f_k$ sourced at $A_1,.....A_k$. We approximate the effect of interference from multiple such sources by simply summing up the amount of resource units consumed at $C$ individually by each such interfering flow $f_i$.
As mentioned in \S\ref{sec:resource-unit}, we simply add up the amount of resource units consumed by individual interfering flows to approximate the cumulative level of interference. 

This is a very close approximation when $\delta(f_i,C) = 1$ (e.g. the duration for which $C$ must back off due to interference will add up for different sources of interference).

In cases where $\delta(f_i,C) < 1$ (i.e. interference effect is captured by degradation in channel quality), the additive model continues to provide a close approximation when $\sum (r(f_i, A_i)) \leq 1$, where we can assume that interference from different sources impacts different portions of the channel (e.g. at different times).

When $\delta(f_i,C) < 1$ and $\sum (r(f_i, A_i)) > 1$, our additive model over-estimates the effects of interference on the portions of channel bandwidth that experience simultaneous interference from multiple sources. In particular, the impact of simultaneous interference from multiple sources on datarate can be approximated as:
\begin{subequations}
\begin{align}
datarate &= BWlog \left( \frac{ss(d)}{I_1 + I_2 + ... I_n} \right) \\
&= BW(log(ss(d)) - log(\sum(I_i))
\end{align}
\end{subequations}
Here, $I_i$ denotes the strength of the interfering signal. 

Our additive model will instead estimate the datarate associated with the leftover channel-capacity (for the portion of channel that experiences simultaneous interference) as: 
\begin{subequations}
\begin{align}
datarate &= BW(log(ss(d)) - \sum(log(I_i))
\end{align}
\end{subequations}
This directly follows from the reasoning about a single interference source detailed above. 

Given positive interference signals (in dB), $I_i > 1$ and $\sum(log(I_i)) > log(\sum(I_i))$, resulting in over-estimated impact of interference. This overestimate is restricted to portion of channel bandwidth that experience simultaneous interference (roughly corresponding to $(\sum r(f_i, A_i) - 1)$ resource units). Moreover, its restricted to settings where impact of interference is smaller to begin with $\delta(f_i,C) < 1$.

\section{More Measurement Results}
\label{appendix:more_measurements}

We complete our analysis from section \ref{sec:measurements} here by looking at what factors contribute to variations in throughput.

\subsection{Trends Over Weeks}
\begin{figure}
\centering
\captionsetup{justification=centering}
\captionsetup[subfigure]{justification=centering}
\centering
\subfloat[Trends Over Months]{\includegraphics[width=0.23\textwidth]{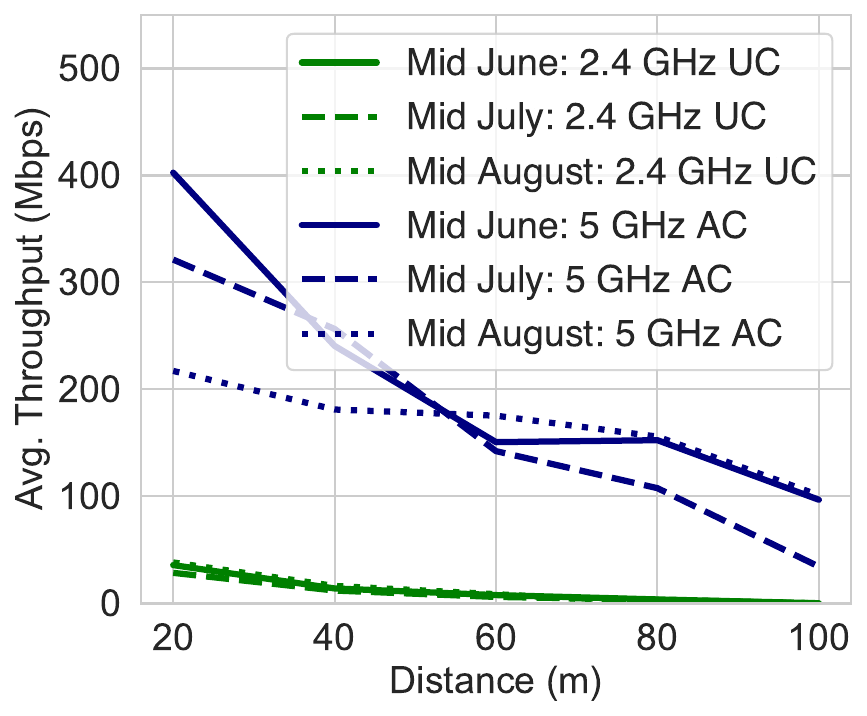} \label{fig:over-months}}
\subfloat[Spatial \& Temporal Variations]{\includegraphics[width=0.23\textwidth]{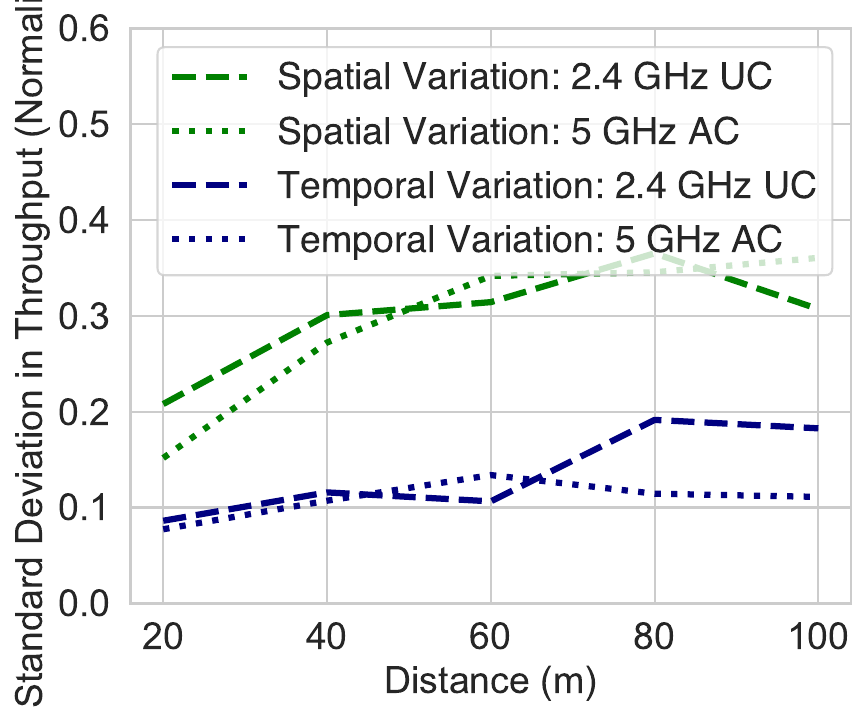} \label{fig:spatial-temporal}}
\vspace{-10pt}
\caption{Variations in throughput over months, and standard deviation in throughput due to temporal and spatial reasons. UC = Under the Canopy, AC = Above the Canopy}
\vspace{-10pt}
\label{fig:variations}
\end{figure}

We next study how WiFi throughput varies over the growing season.
In Figure \ref{fig:over-months}, we show snapshots from three different weeks during three different months for 5GHz above-canopy and 2.4GHz under-canopy throughput. We find that the overall trends remained consistent across the growing season with some overall variations owing to factors like crop age and weather.

\begin{figure}
    \centering
    \includegraphics[width=0.47\textwidth]{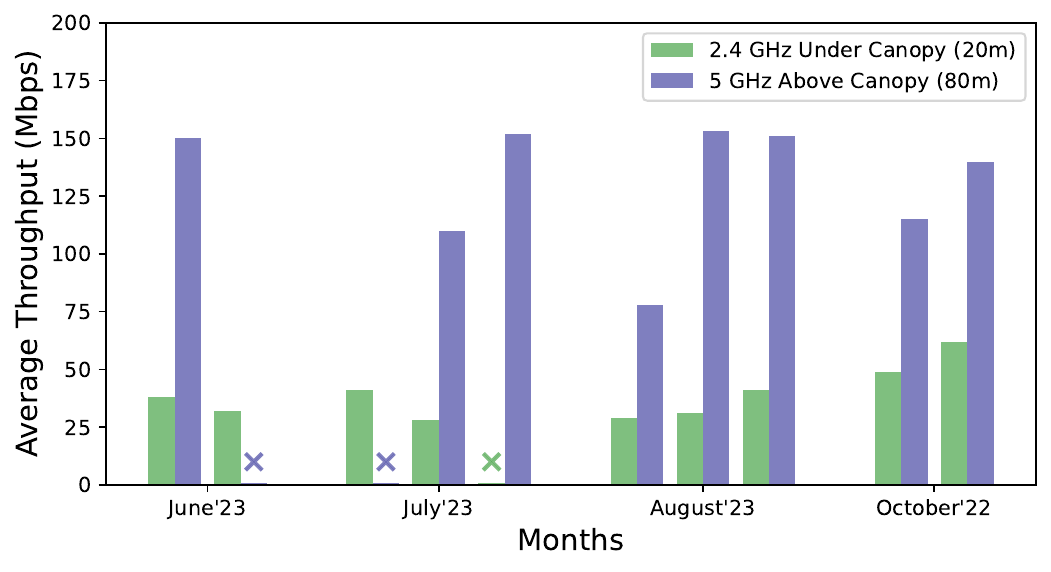}\vspace{-0.2in}
    \caption{Throughput over different weeks of the season. 
    }\vspace{-0.15in}
    \label{fig:trends_over_months}
\end{figure}

Figure \ref{fig:trends_over_months} shows the average throughput for two settings of interest: 2.4 GHz under the canopy at shorter distances (20 m) and 5 GHz above the canopy at longer distances (80 m). Some of the data points are missing since we could not get clean measurements either due to interference from external sources (missing 2.4 GHz data point in July) or because crops had outgrown our mounts for above canopy measurements (missing 5 GHz data point). While figure \ref{fig:over-months} shows that overall trends with respect to distance are consistent across all measurements, these results also highlight that there are still variations over longer time periods. This is likely due to different moisture levels in the crops and environment owing to either recent weather or the age of the crops. For example, under-canopy throughput gets better as crops start to age and dry out.

\subsection{Spatial and Temporal Variations}
\label{sec:variations}

We next report whether there were any significant variations in throughput (i) over time when the transmitting device is stationary, or (ii) when the device moves by a few centimeters (figure \ref{fig:spatial-temporal}). 

\paragraphb{No Significant temporal variations} Due to lack of interference from external sources, we observed fairly steady throughput over the 30s duration of each experiment run, with standard deviations around 10\%. 

\paragraphb{Spatial variations} In contrast, moving an under-canopy device by a few centimeters had more significant impact on throughput (resulting in a standard deviation of around 30 \%. This also indicates scope for improving network performance by adjusting the position of a mobile sender~\cite{dhekne2018if} or mounting the WiFi antennas on a mechanical base that can be maneuvered to find a ``good'' spot (we leave detailed exploration of this to future work).


\section{Multihop Validation}
\label{appendix:multihop}
\begin{figure}
    \centering
    \includegraphics[width=0.35\textwidth]{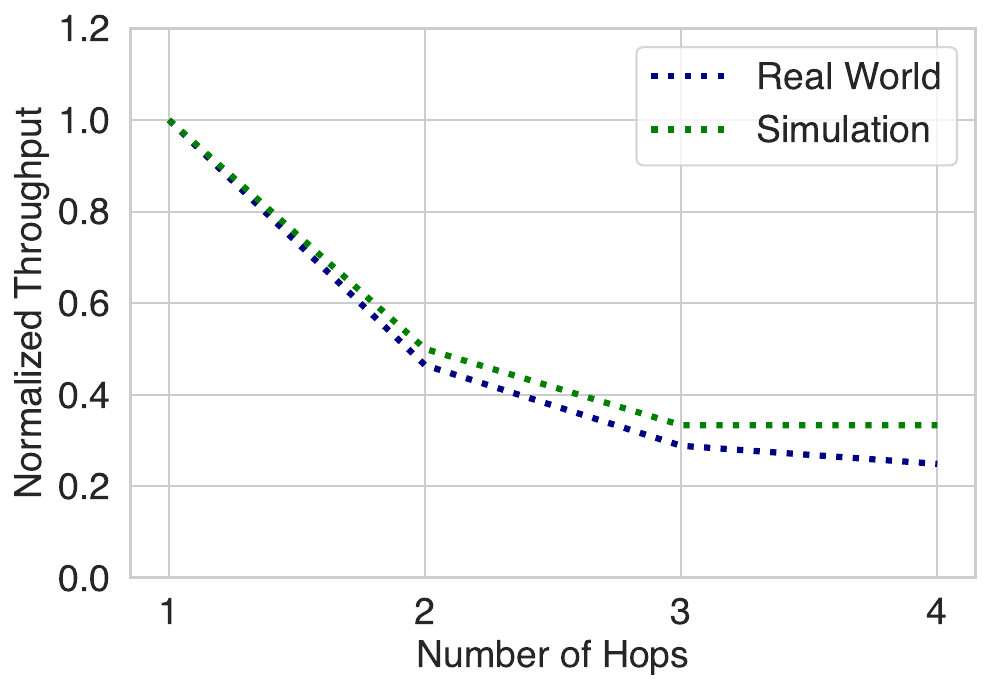}\vspace{-0.2in}
    \caption{Comparing modeling of multihop in our simulator with real-world measurements.
    }\vspace{-0.15in}
    \label{fig:multihop}
\end{figure}

We connected up to 5 routers serially and placed them apart to form up to 4 hops and measured throughput in each scenario. The same experiment was repeated in the simulator. In both cases, the throughput degrades in a similar fashion as shown in figure \ref{fig:multihop}.

\section{Channel-Switching Overheads} 
\label{sec:channel-switching}
We conduct a small-scale experiment to evaluate the feasibility and overhead of remotely configuring WiFi routers and devices by sending control messages over LoRA to change channels and install routes (as proposed for \farmernetes' TM).  
Our setup included 2 LoRA transmitters (WaveShare SX1262 868M) as a sender and receiver, each attached to two separate Raspberry Pi 4B. Sending Raspberry Pi runs the controller while receiving Raspberry Pi is attached to a Wifi router over Ethernet. Upon receiving the control message, the receiving Raspberry Pi makes changes to the router over a pre-connected SSH connection. We measure the overhead of switching the channel of a router in this setup. The message takes about 1.3 seconds to be propagated between controller and the router. Once the router switches its channel, the devices connected to that router disconnect, they scan the WiFi again and reconnect. We measured this overhead by pinging a device connected to the router. Across runs, devices becomes unavailable for ~5.2 seconds during channel switching. 


\end{document}